\documentclass[aip, jap, amsmath, amssymb, reprint]{revtex4-2}

\usepackage{graphicx}
\usepackage{dcolumn}
\usepackage{bm}
\usepackage[utf8]{inputenc}
\usepackage[T1]{fontenc}
\usepackage{mathptmx}
\usepackage{etoolbox}
\usepackage{url}
\usepackage{color}
\usepackage{xcolor}
\usepackage{hyperref}

\begin{document}

\preprint{AIP/123-QED}

\title{Characterization and optimization of high-efficiency crystalline silicon solar cells}

\author{V.P.~Kostylyov}
\affiliation{V. Lashkaryov Institute of Semiconductor Physics, NAS of Ukraine, 41 prospect Nauky, 03028 Kyiv, Ukraine }

\author{A.V.~Sachenko}
\affiliation{V. Lashkaryov Institute of Semiconductor Physics, NAS of Ukraine, 41 prospect Nauky, 03028 Kyiv, Ukraine }

\author{M.~Evstigneev}
\affiliation{Department of Physics and Physical Oceanography, Memorial University of Newfoundland, St. John's, NL, A1B 3X7 Canada}

\author{I.O.~Sokolovskyi}
\affiliation{V. Lashkaryov Institute of Semiconductor Physics, NAS of Ukraine, 41 prospect Nauky, 03028 Kyiv, Ukraine }

\author{A.I.~Shkrebtii}
\affiliation{Faculty of Science, Ontario Tech University, 2000 Simcoe Street North, Oshawa, ON, L1G 0C5 Canada}


\begin{abstract}
Since the photoconversion efficiency $\eta$ of the silicon-based solar cells (SCs) under laboratory conditions is approaching the theoretical fundamental limit, further improvement of their performance requires theoretical modeling and/or numerical simulation to optimize the SCs parameters and design. The existing numerical approaches to modeling and optimization of the key parameters of high-efficiency solar cells based on monocrystalline silicon (c-Si), the dominant material in photovoltaics, are described. It is shown that, in addition to the four usually considered recombination processes, namely, Shockley-Read-Hall, surface, radiative, and band-to-band Auger recombination mechanisms, the non-radiative exciton Auger recombination and recombination in the space charge region (SCR) have to be included.  To develop the analytical SC characterization
formalism, we proposed a simple expression to model the wavelength-dependent external quantum efficiency ($EQE$) of the photocurrent near the absorption edge. Based on this parameterization, the theory developed allows for calculating and optimizing the base thickness-dependent short-circuit current, the open-circuit voltage, and the SC photoconversion efficiency. We proved that the approach to optimize the solar cell parameters, especially its thickness and the base doping level, is accurate and demonstrated for the two Si solar cells reported in the literature, one with an efficiency of 26.7\,\% and the other with the record efficiency of 26.81\,\%. It is shown that the formalism developed allows further optimization of the solar cell thickness and doping level, thus increasing the SC efficiency to an even higher value.
\end{abstract}

\maketitle

\section{Introduction}
Enormous progress has been achieved in the field of photovoltaics since the invention of the first practical silicon solar cell (SC), as published in the Journal of Applied Physics 70 years ago \cite{Chapin54}.  This breaking applied research has initiated a new field of photovoltaic solar energy conversion and transformed it into a major component of the renewable energy sector. More than 60 years ago, Shockley and Queisser in their seminal paper \cite{Shockley61} laid the theoretical foundations for the fundamentals of solar cell operation and related photoconversion efficiency ${\eta}$, see also \cite{Swanson2005, Andreani19, Markvart22,  Ballif22}. After the pioneering research by Shockley and Quiesser \cite{Shockley61}, the theoretical limit of photoconversion efficiency under the AM1.5G irradiation conditions for single-junction Si SCs has been extensively studied by numerous groups \cite{Tiedje84, Green84, Kerr03, Richter13, Schafer18, Sachenko20, Niewelt22, Veith-Wolf18, Engelbrecht21, Evstigneev23}, all reporting the values around 29.5\,\% up to a few tenths of a per cent. 

Even though various strategies to overcome this limit in photovoltaics
have been proposed (see, e.g., Ref.~\onlinecite{Lee24} and references therein), crystalline
silicon (c-Si) based SCs remain the most practical, cost-effective, stable, and
durable component for the photoconversion of solar energy into electricity for
terrestrial applications. As a result, the past two decades have witnessed a remarkable surge in silicon solar cells' efficiency in laboratory \cite{Lin23} and industrial \cite{Ru2024} settings. Notably, researchers and technologists at LONGi have achieved a significant milestone by developing a large-area SC sample with a record-breaking photoconversion efficiency of 26.81\,\% \cite{Lin23}, surpassing the previous record of 26.7\,\% \cite{Taguchi14, Yamamoto18} by 0.1\,\%. This achievement underscores the formidable challenge of further enhancing photoconversion efficiency and suggests that we are approaching the practical limit for silicon SCs with either homo- or heterojunctions. Furthermore, the LONGi group has very recently reported a further whopping efficiency increase to  27.3 \%   \cite{Green24}. 

The technological advancements can be attributed primarily to the improved methods of material growth and manufacturing of low-defect, low-contamination wafers, with photoexited carrier lifetimes of the order of tens of milliseconds. Additionally, there have been significant developments in effective surface passivation techniques and interface engineering \cite{Nakamura14, Masuko14, Feldmann14, Adachi15, Hermle20, Schmidt18}. These developments have led to a notable increase in the measured effective lifetime of charge carriers, while simultaneously reducing the contribution of removable (extrinsic) recombination mechanisms compared to non-removable (intrinsic) ones. This trend is supported by recent research focusing on refining the parameterization of the intrinsic Auger and radiative recombination mechanisms \cite{Niewelt22, Veith-Wolf18, Black22}, which further highlights the critical importance of the accurate description of intrinsic recombination in silicon SCs close to the theoretical efficiency limit. As efforts in improving manufacturing processes and exploring new generation solar cell concepts, such as POLO \cite{Haase17}/TOPCon \cite{Richter17}, continue to evolve, the emphasis on understanding and mitigating the recombination mechanisms in silicon will undoubtedly persist.

The necessity for an accurate description of intrinsic recombination in silicon arises from a broader requirement for the precise characterization of all recombination channels affecting non-equilibrium (excess) charge carriers, both intrinsic and extrinsic, within crystalline silicon-based solar cells. To address this challenge, the development of theoretical principles for the comprehensive characterization of silicon SCs and their practical application becomes essential.

Currently, many tools have been developed to perform SC modeling \cite{Altermatt11, Sugiura23, Richter21, Andreani19}. These tools and software products have shown their effectiveness in modeling and optimizing single-junction silicon SCs with homo- and heterojunction. However, their common limitations are that they all consider four principal
recombination mechanisms in silicon: Shockley-Read-Hall (SRH) recombination, Auger recombination, radiative recombination, and surface recombination. This paper presents an argument that two additional recombination channels need to be included to correctly describe the dependence of the effective lifetime on the excess concentration of charge carriers and the SC output power on the applied voltage: they are recombination in the space-charge region (SCR) and the trap-assisted exciton Auger recombination. 

The latter exciton recombination mechanism \cite{Hangleiter87, Hangleiter88, Szmytkowski20} appears, because the spatial localization of an electron and a hole in an exciton significantly increases the probability of the Auger recombination. In such a process, either an electron or a hole is captured by the trap, while the energy and momentum released are carried away by the other quasiparticle that formed an exciton. This recombination process acts in parallel with the SRH recombination; as a result, the effective carrier lifetime is shorter than the ``pure'' SRH lifetime by a factor that increases with the doping level \cite{Sachenko16, Sachenko24}.

The next feature of our approach is the use of a simple analytical formula \cite{Sachenko20a} to simulate the dependence of the external quantum efficiency on the wavelength and SC thickness near the absorption edge. Although this formula contains only one fit parameter, its comparison with the experimental curves of the short-circuit current and the photoconversion efficiency vs. cell thickness showed excellent agreement \cite{Sachenko20a, Sachenko23}.

To enhance the parameters of next-generation silicon SCs, especially aiming for increased efficiency, a meticulous characterization of these cells stands as a primary requisite. The results reported in our study elucidate that employing six rather than four recombination processes, notably incorporating the recombination rate in the SCR, offers a more comprehensive SC characterization compared to the prevalent practice of utilizing four recombination processes. Our research underscores a crucial finding: it is the recombination within the SCR, rather than surface recombination, that predominantly is responsible for the diminished value of the effective lifetime of nearly all silicon SCs. Thus, prioritizing the investigation of its impact on the attributes of highly efficient silicon SCs is imperative.

\section{External quantum efficiency and short circuit current density}
The external quantum efficiency $EQE(\lambda)$ allows finding the short-circuit current density given the spectral density of the incident photon flux $I(\lambda)$ as
\begin{equation}
J_{SC} = q\,\int d\lambda\,EQE(\lambda)\,I(\lambda)
\label{10}
\end{equation}
where $q$ is the elementary charge. In the actual SCs, the $EQE(\lambda)$ dependence is determined by such factors as the chemical composition and morphology of the surface, presence of a coating with an ITO layer or a grid for current collection, the absorption coefficient of the semiconductor, and other factors.

However, the external quantum efficiency $EQE(\lambda)$ can be modeled with an analytical expression that involves several fit parameters \cite{McIntosh15, Fell16}. We will be using a simpler alternative expression that gives this function for SCs of arbitrary base thickness and uses just two fit parameters \cite{Sachenko20a}. These parameters are determined from a single experimental $EQE(\lambda)$ curve measured for one particular thickness. The idea of this approach is as follows.

The wavelength-dependent external quantum efficiency curve can be divided into two regions: the short-wavelength region $\lambda < 800\,\text{nm}$, which is denoted by the index $s$, in which the external quantum efficiency practically does not depend on the sample thickness $d$, and the long-wavelength region $\lambda > 800\,\text{nm}$, denoted by the index $l$, in which the thickness dependence is present. In the long-wavelength region, the fit expression is
\begin{equation}
EQE_l(\lambda) = \frac{f}{1 + b\,(4n_r^2(\lambda)\,\alpha(\lambda)\,d)^{-1}}
\label{20}
\end{equation}
where the fit parameter $b$ determines the shape of the curve $EQE_l(\lambda)$, and the parameter $f$ is chosen so as to match the values of $EQE_s$ and $EQE_l$ at $\lambda = 800\,\text{nm}$. The non-dimensional parameter $b$ has the physical meaning of the ratio of the photon mean path length $4n_r^2d$ in an SC with ideal Lambertian surfaces to its actual mean photon path length. It depends on the surface texturing and the base thickness. For an ideally diffusive surface, this parameter has the value $b = 1$. Note that for the special choice $f = b = 1$, the expression (\ref{20}) turns into the well-known absorptance formula \cite{Tiedje84}.

In the short-wavelength region $\lambda < 800\,\text{nm}$, the experimental $EQE_s(\lambda)$ curves do not depend on the thickness, but are determined only by the losses due to reflection, shading, and absorption of light outside the SC base region. The digitized $EQE_s(\lambda)$ curve obtained for one particular thickness is assumed to apply to samples of any thickness.

\section{ Lifetime of the photoexcited carriers in silicon}
\label{section3}
The total  lifetime of the photoexcited carriers in silicon SCs is defined by the intrinsic and extrinsic recombination mechanisms:
\begin{equation}
\tau_{eff}^{-1} = \tau_{intr}^{-1} + \tau_{extr}^{-1}\ ,
\end{equation}
where the intrinsic lifetime is formed by the radiative and Auger contributions,
\begin{equation}
\tau_{intr}^{-1} = \tau_{r}^{-1} + \tau_{A}^{-1}\ .
\end{equation}
The extrinsic lifetime 
\begin{equation}
\tau_{extr}^{-1} = \tau_{SRH}^{-1} + \tau_{S}^{-1} + \tau_{exc}^{-1} + \tau_{SCR}^{-1}
\end{equation}
is formed by the Shockley-Read-Hall and surface recombination mechanisms (the first two terms in the above expression), as well as the exciton trap-assisted Auger and space charge region recombination channels (the third and the fourth terms). We note that in the majority of the existing approaches, the last two recombination channels are usually left out.

The non-radiative exciton Auger recombination mediated by a deep recombination level was first observed by Fossum \cite{Fossum76}. Its model was developed later  by Hangleiter \cite{Hangleiter87, Hangleiter88}, who related the relevant lifetime to the SRH lifetime and the doping level $n_0$ as
\begin{equation}
\tau_{exc} = \tau_{SRH}\,\frac{n_x}{n_0}\ ,
\end{equation}
where the characteristic concentration was originally assigned a value $n_x = 7.1\cdot10^{15}\,\text{cm}^{-3}$. This value was later refined\cite{Sachenko16} to $8.2\cdot10^{15}\,\text{cm}^{-3}$ after analyzing a number of experimental works, in which the existence of this recombination mechanism is confirmed. This mechanism is also described in detail in the monograph \cite{Abakumov91} and in the second appendix of the recent paper \cite{Sachenko24}.

We note that, in fact, it is not the SRH lifetime that is determined from the experiment, but the effective lifetime
\begin{equation}
\tau_{SRH, eff} = \frac{\tau_{SRH}}{1 + n_0/n_x}\ .
\end{equation}
The experimentally determined effective SRH lifetime is always shorter than the ``bare'' SRH lifetime; the difference between the two increases with the level of base doping.

As for the recombination in the SCR, it will be shown below that without taking it into account, theoretical agreement with the experiment cannot be obtained for a number of characteristics of silicon SCs, especially for illuminated $I-V$ curves and for the relation between the SC output power and the applied voltage, as well as for the effective lifetime vs. the excess concentration curves.

The radiative lifetime is given by the expression\cite{Richter13}
\begin{equation}
\tau_r^{-1} = B(1 - P_{PR})(n_0 + \Delta n)\ ,
\label{80a}
\end{equation}
where $P_{PR}$ is the probability of photon recycling and $B$ the radiative recombination coefficient given by \cite{Wurfel82}
\begin{align}
&B = \int_0^\infty dE\,B(E)\ ,\nonumber\\
&B(E) = \frac{8\pi}{h^3}\,\alpha_{bb}(E)\,\left(\frac{n_r(E)\,E}{c_0\,n_{i, eff}}\right)^2\,e^{-E/kT}\ .
\end{align}
Here, $c_0$ is the speed of light in vacuum, $h$ is Planck's constant, $n_r(E)$ is the refractive index, and $\alpha_{bb}(E)$ is the absorption coefficient for the photons of energy $E$ that result in creation of an electron-hole pair, and 
\begin{equation}
n_{i, eff}(V) = n_i\,e^{\Delta E_g(V)/(2kT)}
\end{equation}
is the effective intrinsic concentration corrected with respect to the bandgap narrowing effect by the amount $\Delta E_g$. \cite{Schenk98} The photon recycling probability is given by
\begin{equation}
P_{PR} = B^{-1}\,\int_0^\infty dE\,B(E)\,A_{bb}(E)\ ,
\end{equation}
where the absorptance is
\begin{equation}
A_{bb} = \frac{\alpha_{bb}}{\alpha_{bb} + \alpha_{FCA} + b/(4n_r^2 d)}\ .
\label{110}
\end{equation}
The second term in the denominator is the coefficient of absorption by the free charge carriers \cite{Rudiger13}. This expression differs from the Tiedje-Yablonovich formula \cite{Tiedje84} by the parameter $b > 1$.

Auger recombination time is given by the general expression
\begin{equation}
\tau_{Auger} = \frac{\Delta n}{C_eg_e(n^2p - n_0^2p_0^2) + C_hg_h(np^2 - n_0p_0^2)}\ ,
\label{120a}
\end{equation}
where $n$ and $p$ are the electron and hole concentrations, $n_0$ and $p_0$ are their respective values in thermal equilibrium, $C_{e, h}$ are the Auger coefficients, and the non-dimensional factors $g_{e, h}$ account for the enhancement of the Auger process due to the Coulomb screening. Several expressions have been proposed for these parameters \cite{Richter12, Niewelt22, Veith-Wolf18}. They will be compared in the context of the SC modeling below.

The SRH lifetime depends on the doping and excitation level in an n-type semiconductor as
\begin{equation}
\tau_{SRH} = \frac{\tau_{p0}(n_0 + n_1 + \Delta n) + \tau_{n0}(p_1 + \Delta n)}{n_0 + \Delta n}\ ,
\end{equation}
where the characteristic times $\tau_{p0} = 1/(V_p\sigma_pN_t)$ and $\tau_{n0} = 1/(V_n\sigma_nN_t)$ depend on the hole and electron thermal velocities, $V_p$ and $V_n$, and the respective capture cross-sections, $\sigma_p$ and $\sigma_n$, by a trap of concentration $N_t$ and energy $E_t$. Finally, the concentrations $n_1$ and $p_1$ are the electron and hole concentrations when $E_F = E_t$. Depending on the excess concentration $\Delta n$ of the electron-hole pairs, the SRH lifetime changes between the low-injection and the high-injection extreme values.

The non-radiative exciton Auger lifetime is \cite{Sachenko16}
\begin{equation}
\tau_{exc} = \tau_{SRH}\,\frac{n_x}{n_0 + \Delta n}\ .
\label{140a}
\end{equation}
The surface recombination time is
\begin{equation}
\tau_S = d/S\ ,
\end{equation}
where $S$ is the net recombination velocity on the front and the rear surface. It depends on the doping and excitation levels according to
\begin{equation}
S = S_0\,\left(\frac{n_0}{n_p}\right)^m\,\left(1 + \frac{\Delta n}{n_0}\right)^r\ .
\end{equation}
In this expression, $S_0(n_0/n_P)^m$ is the surface recombination velocity at a low excitation level, $n_p$ is the doping level at which $S = S_0$ for $\Delta n = 0$, and the exponents $m$ and $r$ are both of the order of 1.

Similarly, the SCR recombination time is related to the respective recombination velocity by
\begin{equation}
\tau_{SCR} = d/S_{SCR}\ .
\end{equation}

In unprocessed silicon wafers intended for SC fabrication and used in the experimental measurements of the lifetime, the $\tau_{eff}(\Delta n)$ curves are known to saturate \cite{Richter12} at $\Delta n < 10^{15}\,\text{cm}^{-3}$, whereas the effective lifetime in Si SCs has a maximum at some excess carrier concentration \cite{Richter17, Ru2024}. This difference can be attributed to the fact that in the c-Si samples used for the measurements, there is no p-n junction, hence the band bending on both surfaces of the Si wafer is symmetrical and slightly depleting. In this case, the SCR recombination is practically absent. On the other hand, after the SC has been fabricated, a p-n junction is formed near one of the wafer surfaces with significant recombination in the SCR at voltages up to the maximum power collection voltage $V = V_m$ with $V_m = 0.55 - 0.65$\,V. This is explained by the fact that the SRH lifetime in the SCR is always significantly lower than the SRH lifetime in the quasineutral region \cite{Sachenko24, Dauwe04, Veith-Wolf18a}. In turn, this is caused by the fact that the concentration of deep impurities is usually much higher in the SCR than in the quasineutral region.

A similar situation occurs when silicon is passivated with the dielectric layers of SiN$_x$ or Al$_2$O$_3$, and a significant charge is built into the dielectric \cite{Dauwe04, Veith-Wolf18a}. In the case of a SiN$_x$ passivation layer, this charge is positive, and in the case of Al$_2$O$_3$, it is negative. Then, near the surface of p-Si in the first case and n-Si in the second case, the conductivity inversion condition occurs and the recombination in the SCR becomes significant. This was confirmed\cite{Sachenko24, Dauwe04, Veith-Wolf18a} primarily by the fact that the dependence of the effective lifetime on excess concentration observed a maximum and a decrease in $\tau_{eff}$ in the region of $\Delta n < 10^{15}\,\text{cm}^{-3}$. The lifetimes in SCR were significantly lower than in the neutral bulk of the base and were on the order of microseconds \cite{Sachenko24, Dauwe04, Veith-Wolf18a}.

The SCR recombination velocity was calculated as
\begin{align}
S_{SCR}(\Delta n) &= \int_0^w dx\,(n_0 + \Delta n)\,\Big[(n_0+\Delta n)\,e^{y(x)} + n_i\,e^{E_t/kT}\nonumber\\
&+ b_r\Big((p_0 + \Delta n)\,e^{-y(x)} + n_i\,e^{-E_t/kT}\Big)\Big]^{-1}\,\tau_{SCR}^{-1}(x)\ .
\label{170}
\end{align}
Here, $b_r = C_p/C_n = V_{pT}\sigma_p/(V_{nT}\sigma_n)$ is the ratio of the hole and electron capture coefficients, expressed as products of the respective thermal velocities $V_{(p, n)T}$ by the capture cross-sections $\sigma_{p, n}$, $\tau_{SCR} = (V_{pT}\sigma_p\,N_t^*)^{-1}$ is the hole lifetime in the SCR, $N_t^*$ is the concentration of deep impurities in the SCR, and $n_0$ and $p_0$ are the equilibrium electron and hole concentrations, $y(x)$ is the electric potential divided by the thermal voltage $kT/q$, $E_t$ is the energy of the deep impurity measured from the middle of the band gap, and $w$ is the SCR width found from
\begin{equation}
w = \int_{y_0}^{y_w}dx\,\frac{L_D}{\sqrt{\left(1 + \frac{\Delta n}{n_0}\right)\,(e^{y}-1) - y + \frac{\Delta n}{n_0}(e^{-y} - 1)}}\ ,
\end{equation}
where $y_0$ is the non-dimensional potential at the boundary between the n- and p-regions, $y_w$ is the non-dimensional potential at the boundary between the SCR and the quasineutral regions, and $L_D = \sqrt{\epsilon_0\,\epsilon_{Si}\,kT/(2q^2n_0)}$ is Debye screening length. In the calculations, we took $y_w = -0.1$.

When $\tau_{SCR}(x) = \text{const}$, the integral (\ref{170}) can be simplified as follows. Changing the integration variable from the position $x$ to the non-dimensional potential $y$, we have
\begin{align}
S_{SCR}(\Delta n) &=  \int_{y_w}^{y_0} dy\,(n_0 + \Delta n)\,F\,\Big[(n_0 + \Delta n)\,e^y + n_i\,e^{E_t/kT} \nonumber\\
&+ b_r\Big((p_0 + \Delta n)\,e^{-y} + n_i\,e^{-E_t/kT}\Big)\Big]^{-1}\,\tau_{SCR}^{-1}\ ,
\label{190}
\end{align}
where
\begin{equation}
F = \frac{L_D}{\sqrt{\left(1 + \frac{\Delta n}{n_0}\right)(e^y - 1) + y_m + \frac{p_0 + \Delta n}{n_0}(e^{-y} - 1)}}\ .
\end{equation}

In the following, we will analyze the case when the deep impurity level $E_t$ is in the middle of the bandgap; then, the terms that multiply $n_i$ in (\ref{190}) can be neglected.

The non-dimensional potential $y(x)$ can be found from Poisson's equation, whose solution reads
\begin{equation}
x = L_D\int_{y_0}^y \frac{dy'}{\sqrt{\left(1 + \frac{\Delta n}{n_0}\right)(e^{y'} - 1) - y' + \frac{\Delta n}{n_0}(e^{-y'} - 1)}}\ .
\end{equation}
The non-dimensional potential $y_0$ at $x = 0$ is found from the neutrality equation 
\begin{align}
q\,N = &\sqrt{2kT\,\epsilon_0\epsilon_\text{Si}}\nonumber\\
&\sqrt{(n_0 + \Delta n)(e^{y_0} - 1) - n_0y_0 + \Delta n(e^{-y_0} - 1)}\ ,
\end{align}
where $qN$ is the surface charge density of acceptors in the pn junction \cite{Sachenko24}.

When the voltage applied to the pn-junction increases, or when the SC is illuminated, the value of $y_0$ decreases, and so does the SCR width $w$. That is, part of the SCR becomes neutral. In the region that has become neutral, recombination also occurs, the rate of which is equal to
\begin{equation}
S_{SCR-n} = \frac{w(\Delta n = 0) - w(\Delta n)}{\tau_{SCR}}\,\frac{n_0 + \Delta n}{n_0 + \Delta n + b_r\Delta n}
\label{23}
\end{equation}
For highly efficient silicon SCs, the value of $S_{SCR-n}$ is of a similar order of magnitude as $S_{SCR}$ at the points of maximum power and open circuit. This is because the lifetimes in the SCR are much shorter than the lifetimes in the neutral base. Therefore, when calculating the SCR recombination velocity, one should use the expression in which both components are taken into account:
\begin{equation}
S_{SCR-tot} = S_{SCR} + S_{SCR-n}
\label{24}
\end{equation}
In the further analysis, we will use this expression for the total recombination velocity in the SCR.

\section{Expressions for calculating the  SC characteristics}
We summarize in this section the expressions used to calculate the light and dark dependencies and efficiency of the photoconversion. The illuminated $I-V$ curve is calculated with the help of the following expressions:
\begin{align}
&I_L(V) = I_{SC} - I_r(V) + \frac{V + I_LR_S}{R_{SH}}\ ,\nonumber\\
&I_r(V) = qA_{SC}\left(\frac{d}{\tau^b_{eff}} + S_0\left(1 + \frac{\Delta n}{n_0}\right) + S_{SC-tot}\right)\,\Delta n(V)\ .\nonumber\\
&\tau^b_{eff} = \left(\frac{1}{\tau_{SRH}} + \frac{1}{\tau_{exc}} + \frac{1}{\tau_r} + \frac{1}{\tau_{Auger}}\right)^{-1}\ ,\nonumber\\
&\Delta n = -\frac{n_0}{2} + \sqrt{ \frac{n_0^2}{4} + n_i^2(T, \Delta E_g)(e^{q(V + I_LR_s)/kT} - 1)^2}\ .
\label{300}
\end{align}
Here, $I_L(V)$ is the net current, $I_{SC}$ is the short-circuit current, $I_r(V)$ is the recombination (dark) current, $A_{SC}$ is the SC surface area, $R_s$ and $R_{SH}$ are the series and shunt resistance, and $\Delta E_g$ is the band gap narrowing of Si. The lifetimes $\tau_{exc}$, $\tau_r$, and $\tau_{Auger}$ depend on the total concentration $n = n_0 + \Delta n$ of electrons in the base region according toEqs.~(\ref{140a}), (\ref{80a}), and (\ref{120a}), respectively.

These expressions allows to obtain the open-circuit voltage by setting $I_L = 0$ and $\Delta n(V_{OC}) = \Delta n_{OC}$:
\begin{equation}
V_{OC} = \frac{kT}{q}\,\ln\left(1 + \frac{\Delta n_{OC}(n_0 + \Delta n_{OC})}{n_i^2(T)\,e^{\Delta E_g/kT}}\right)\ .
\end{equation}

Multiplying the current $I_L(V)$ by the applied voltage $V$, we obtain the power $P(V)$, and from the condition of maximum $dP/dV = 0$, we find the value of the voltage at the point of maximum power collection $V_m$. By substituting Vm into the first equation (\ref{300}), we obtain the value of the current $I_m$ at maximum power. This allows to calculate the efficiency $\eta$ of photoconversion and the fill factor $FF$ of the $I-V$ curve in the usual way
\begin{equation}
\eta = \frac{I_m\,V_m}{P_S}\ ,\ \ FF = \frac{I_mV_m}{I_{SC}V_{OC}}\ ,
\end{equation}
where $P_S$ is the incident energy flux density under the AM1.5 conditions.

The dark current is given by 
\begin{align}
&I_d(V) = \frac{qA_{SC}d\,\Delta n}{\tau_{eff}(\Delta n)} - \frac{V - I_dR_s}{R_{SH}}\ ,\nonumber\\
&\Delta n = -\frac{n_0}{2} + \sqrt{\frac{n_0^2}{4} + n_i^2(T, \Delta E_g)\left(e^{q(V - I_dR_s)/kT} - 1\right)}\ .
\end{align}
Finally, the short-circuit current is related to the open-circuit voltage by
\begin{align}
&I_{SC}(V) = \frac{qA_{SC}d\,\Delta n_{OC}}{\tau_{eff}(\Delta n_{OC})} - \frac{V_{OC}}{R_{SH}}\ ,\nonumber\\
&\Delta n_{OC} = -\frac{n_0}{2} + \sqrt{\frac{n_0^2}{4} + n_i^2(T, \Delta E_g)\left(e^{q(V_{OC} - I_dR_s)/kT} - 1\right)}\ .
\end{align}
The effective lifetime can be related to the excess concentration either the short-circuit current vs. the open-circuit voltage or using the dark current. Namely,
\begin{equation}
\tau_{eff} = qd\,\frac{\Delta n_{OC}}{J_{SC}} = qd\,\frac{\Delta n(V)}{J_d(V)}
\end{equation}
When $I_{SC}$ vs. $V_{OC}$ is measured, the excess carrier density $\Delta n_{OC} = \Delta n(n_0, V_{OC})$, and the procedure of finding $\tau_{eff}(\Delta n)$ is quite straightforward. On the other hand, when one uses the dark $I-V$ curve, the excess concentration depends on many parameters, namely, $\Delta n = \Delta n(n_0, R_s, V, I_L)$.

\section{Comparison of the experimental results with the theory }

\subsection{Photoelectric characteristics}
We apply the formalism above to (i) the Sanyo HIT SC\cite{Taguchi05, Taguchi08}, (ii) to a commercial SC with a p-n junction manufactured by the SunPower technology (see Ref.~\onlinecite{Sachenko21} for details), (iii) the HIT SC from Ref.~\onlinecite{Yoshikawa17}, and (iv) to the HIT element with the record for efficiency to date from Ref.~\onlinecite{Lin23}.

\begin{figure} [h!]
    \centering
    \includegraphics[width=1\linewidth]{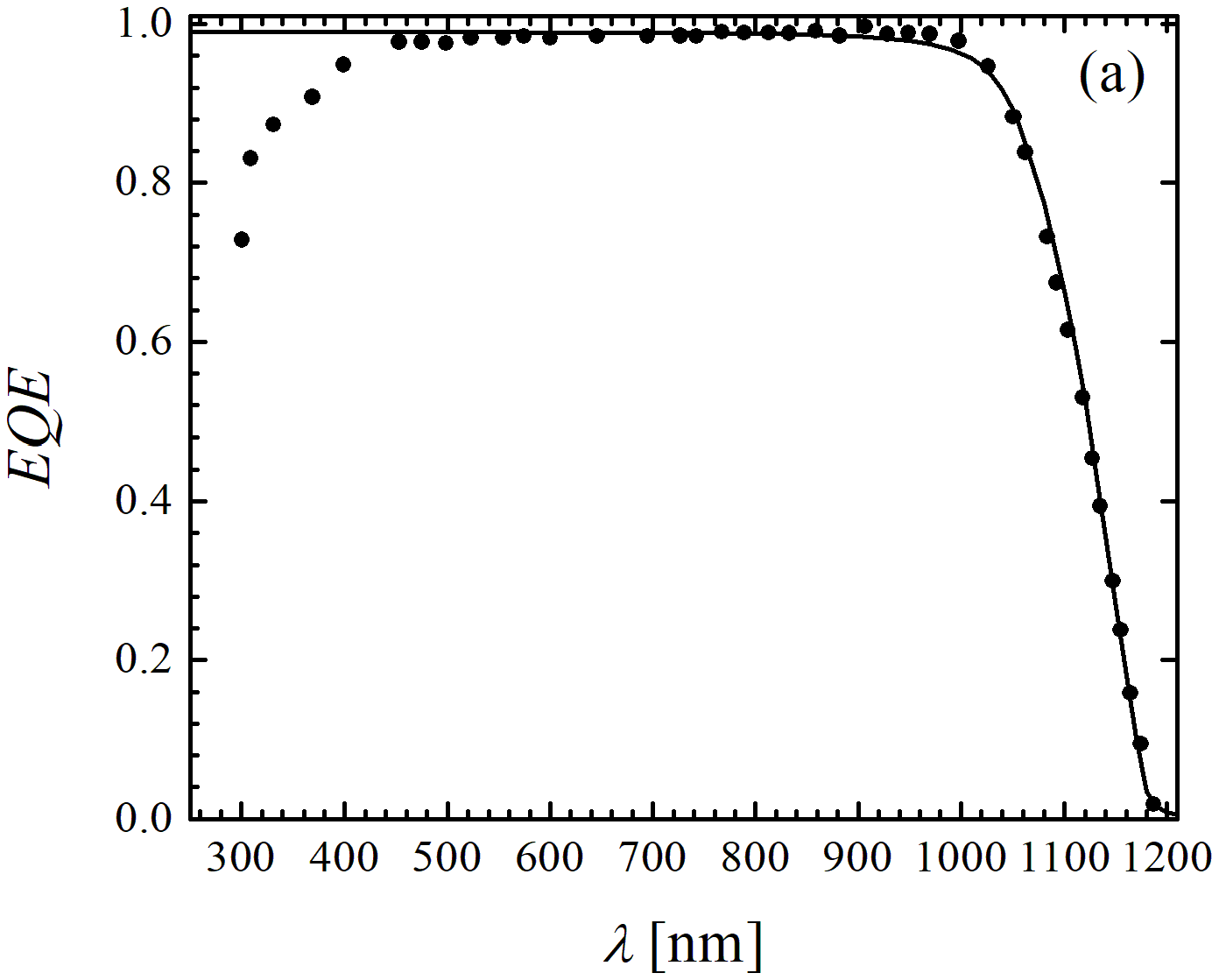}
    \includegraphics[width=1\linewidth]{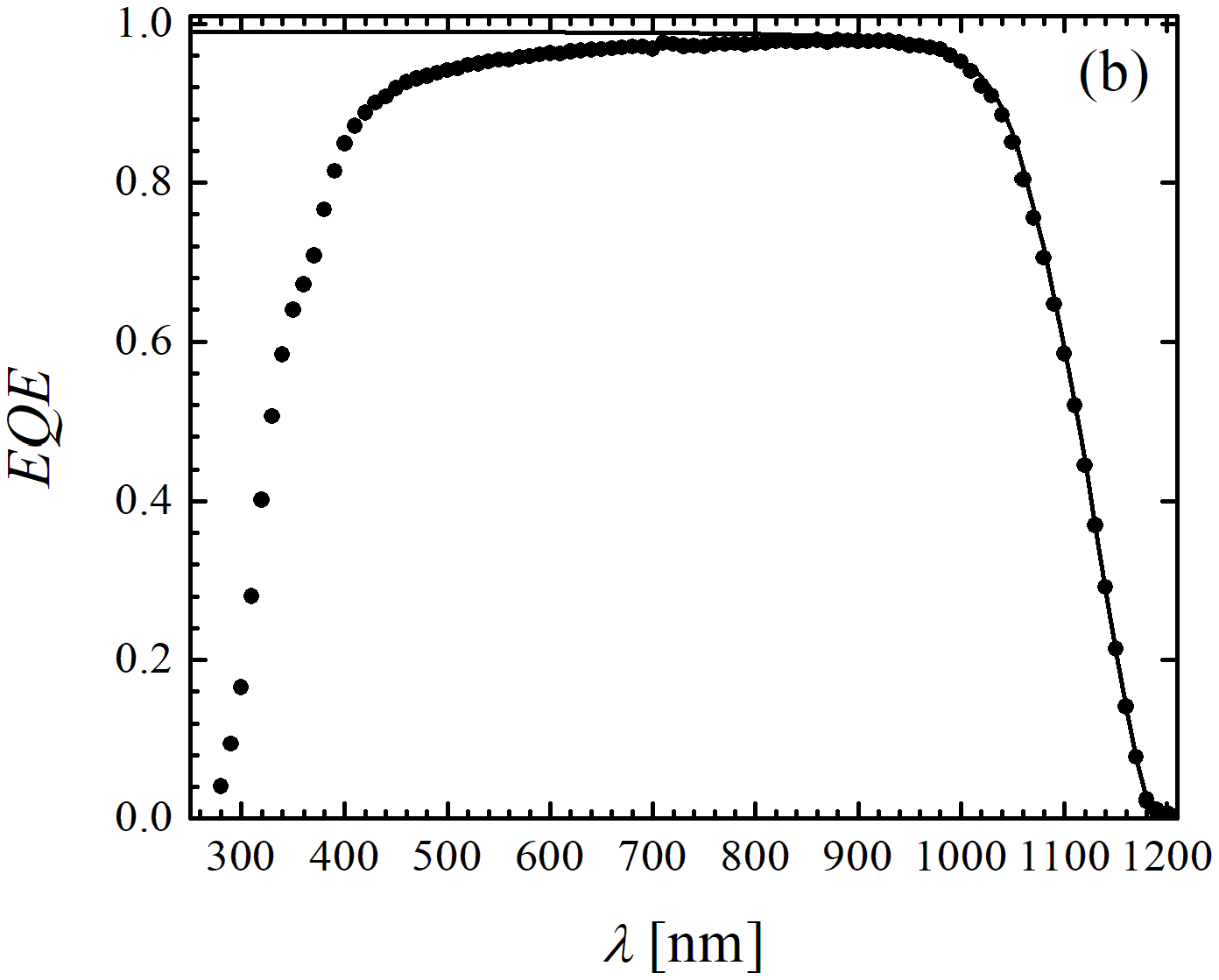}
    \caption{Symbols: Experimental dependence of the external quantum efficiency $EQE(\lambda)$, (a) as measured by Yoshikawa et al. \cite{Yoshikawa17} and (b) Lin et al. \cite{Lin23} (symbols). The theoretical fit (solid lines) of the long-wavelength ($\lambda > 800$\,nm) part of the graph with Eq.~(\ref{20}) yields the parameter $b = 1.8$ in (a) and \textit{b} = 1.6 in (b).}
    \label{fig1}
\end{figure}

In Fig.~\ref{fig1} the experimental  $EQE(\lambda)$ curves, measured (a) by Yoshikawa et al. \cite{Yoshikawa17} and (b) Lin et al. \cite{Lin23}, are compared with the calculated ones. The experimental curves were fitted with Eq.~(\ref{20}) in the long-wavelength ($\lambda > 800$\,nm) part of the spectrum by varying the parameter $b$, which turned out to have the value of 1.8 and 1.6, respectively. Once the parameter $b$ is established, the expressions (\ref{10}) and (\ref{20}) are used to predict how the short-circuit current depends on the base thickness $d$. 

\begin{table}[]
    \centering
    \begin{tabular}{|c|c|c|c|c|c|c|}
    \hline
    SC Ref. & $\eta_\text{exp}$, \% & $\eta_\text{theory}^0$, \% & $\Delta_\eta$ & $\tau_\text{exp}$, ms & $\tau_\text{theory}$, ms & $\Delta_\tau$ \\
    \hline
    \onlinecite{Taguchi05, Taguchi08} & 21.5 & 22.02 & 1.024 & 0.58 & 1.13 & 1.95 \\
\hline \onlinecite{Sachenko21} & 21.8 & 22.07 & 1.012 & 2.82 & 8.83 & 3.13 \\
\hline \onlinecite{Yoshikawa17} and \onlinecite{Sachenko21a} & 26.63 & 26.76 & 1.01 & 7.08 & 13.27 & 1.87 \\
\hline \onlinecite{Lin23} & 26.81 & 27.27 & 1.017 & 5.15 & 18.7 & 3.63 \\
\hline
    \end{tabular}
    \caption{Experimental photoconversion efficiencies and the maximum lifetimes for the SCs from articles indicated in the first column. The theoretical efficiency is obtained without accounting for the SCR recombination, and the lifetimes are found by assuming that the SCR recombination time is the same as the SRH time in the bulk. The parameters $\Delta_\eta$ and $\Delta_\tau$ are the ratio of the theoretical to the experimental efficiency and maximum lifetime respectively.}
    \label{table1}
\end{table}

Let us consider the inaccuracy that may arise if the SCR recombination is neglected in the SC modeling. This analysis is performed for the above-mentioned SCs based on the respective experimental results and the theory developed in this work. Its results are summarized in Table~\ref{table1}, which presents the experimental data for the photoconversion efficiency and the effective lifetimes at the maxima of the $\tau(\Delta n)$ curves found in the literature\cite{Taguchi05, Taguchi08, Sachenko21, Yoshikawa17, Lin23}. Also indicated in this table are the theoretical photoconversion efficiency 
$\eta_\text{theor}^0$ found under the assumption that the SCR recombination is absent, and the maximum lifetimes calculated with the assumption that the SCR lifetime equals the SRH lifetime in the base region. The parameters $\Delta_\eta = \eta_\text{theory}^0/\eta_{exp}$ and $\Delta_\tau = \tau_\text{theory}/\tau_{exp}$ are the ratios of the so obtained theoretical and experimental values.

As can be seen from Table~\ref{table1}, the effect of SCR recombination on the efficiency of the SCs considered here is of the order of one percent. However, the effective lifetimes in the absence of the SCR recombination, that is at low SCR recombination velocities, assumed to be equal to SRH lifetimes in the base region and in the SCR, increase by more than 100\,\%.

The experimental total recombination velocity in the SCR and in that part of the SCR that has become neutral is found from
\begin{align}
S_{SCR-tot}^{exp} =& d\,\big((\tau_{eff}^{exp})^{-1}\nonumber\\
&- \tau_{SRH}^{-1} - \tau_{ex}^{-1} - \tau_S^{-1} - \tau_{Auger }^{-1} -\tau_{r}^{-1}\big)
\end{align}

\begin{figure}
    \centering
    \includegraphics[width=1\linewidth]{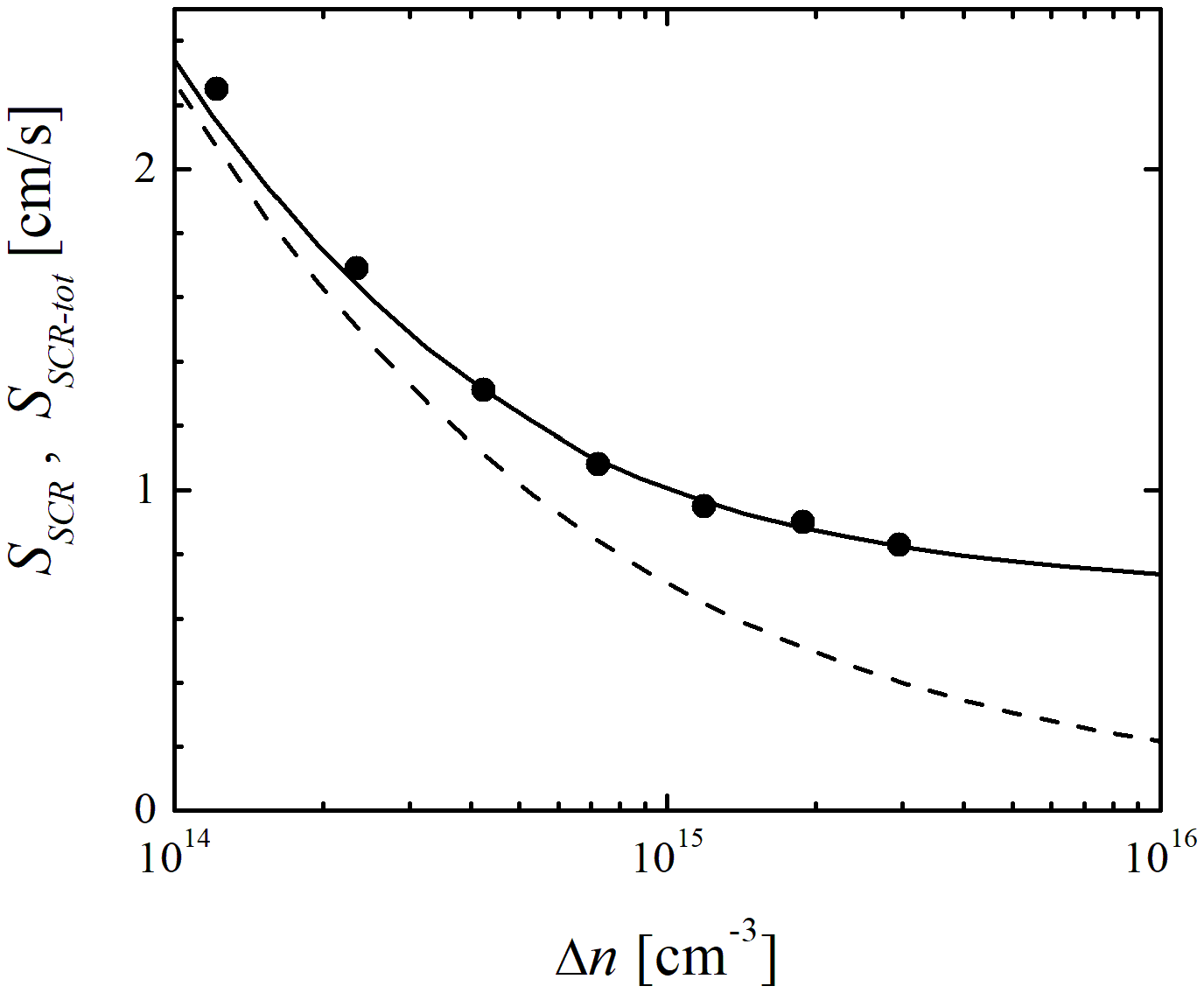}
    \caption{The SCR recombination velocity $S_{SCR}$ (dashed line) and  the total SCR recombination velocity ${S_{SCR-tot}}$ (solid line) vs. the excess concentration $\Delta n$ (black circles), obtained for the SC of Yoshikawa et al. \cite{Yoshikawa17}. Dashed line: theoretical curve (\ref{20}). Solid line: the total surface recombination velocity (\ref{24}).}
    \label{fig2}
\end{figure}

The theoretical SCR recombination velocity is found from expression (\ref{24}).
Fig.~\ref{fig2} shows the experimental and theoretical dependencies of the recombination rate in the SCR on the excess concentration, obtained using the parameters from Yoshikawa et al. \cite{Yoshikawa17}. As can be seen from the figure, the theoretical and experimental dependencies agree well with each other. The figure also shows that at an excess concentration of the order of $3\cdot10^{15}\,\text{cm}^{-3}$, which corresponds to the excess concentration at the point of maximum power, the two contributions to the expression (\ref{24}), namely, the SCR recombination velocity $S_{SCR}$ and the recombination velocity in the region that has become neutral, $S_{SCR-n}$, become comparable.

\begin{figure}
    \centering    \includegraphics[width=1\linewidth]{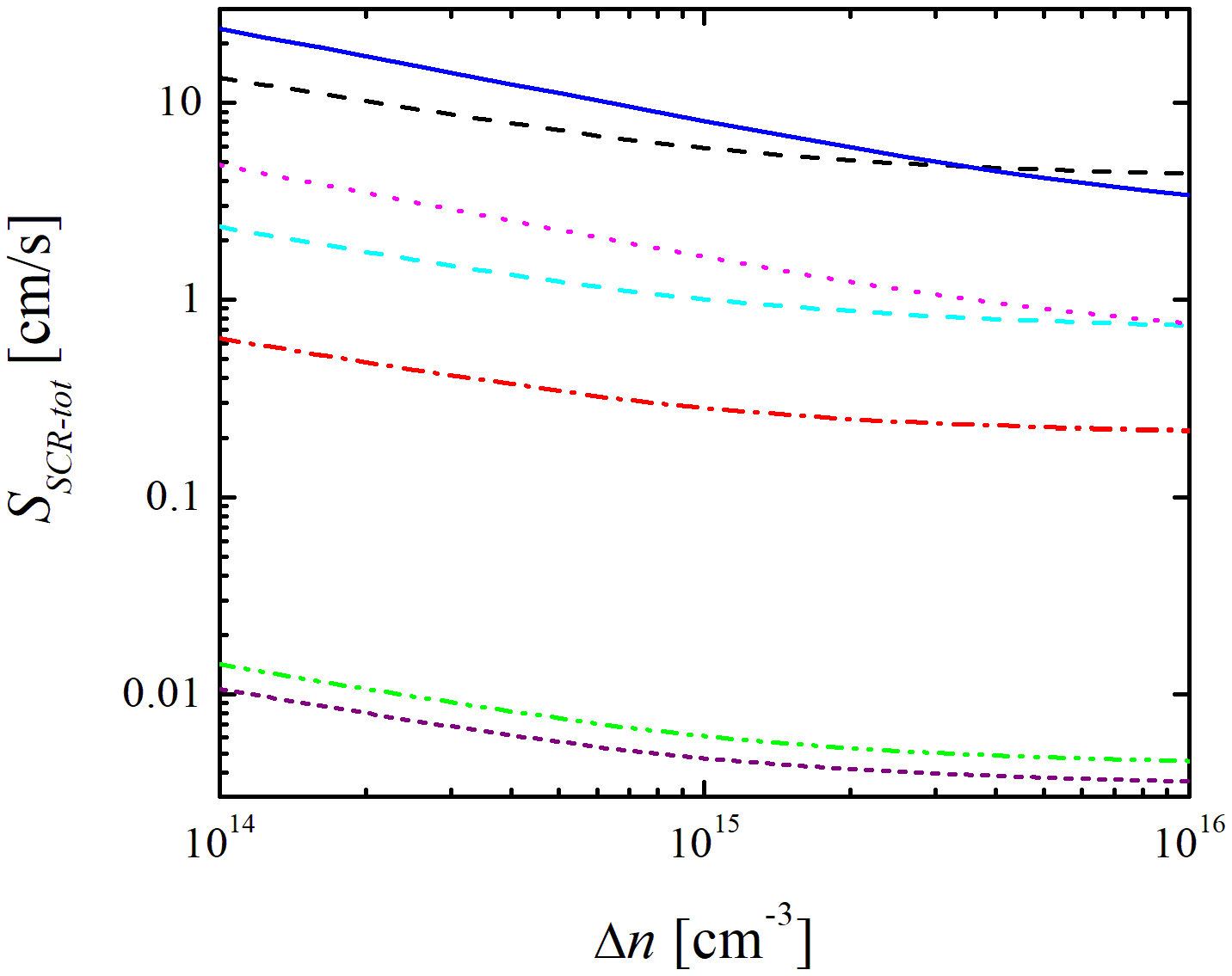}
    \caption{Theoretical total SCR recombination velocity ${S_{SCR-tot}}$ vs. the excess concentration $\Delta n$, plotted using the experimental data (from top to bottom) of Taguchi et al. \cite{Taguchi05}, Sachenko et al. \cite{Sachenko21}, Lin et al. \cite{Lin23}, Yoshikawa et al. \cite{Yoshikawa17}, and Richter et al. \cite{Richter17}. The two lowest theoretical curves are plotted under the assumption that the lifetime in the SCR is equal to the bulk lifetime in the sample discussed in Richter et al. \cite{Richter17} (top) and Yoshikawa et al. \cite{Yoshikawa17} (bottom). }
    \label{fig3}
\end{figure}

Figure~\ref{fig3} shows the theoretical SCR recombination velocity ${S_{SCR-tot} ({\Delta n})}$ curves for all the SCs that appear in Table~\ref{table1}, both the p-n junction-based ones and the HIT elements. In addition, Fig.~\ref{fig3} shows $S_{SCR-tot}(\Delta n)$ curve for the SC described in Ref.~\onlinecite{Richter17}, with the respective calculation performed in Ref.~\onlinecite{Sachenko20a} (red dash-dotted curve). It turned out to have the smallest value of all the SCs analyzed here.

Three curves in Fig.~\ref{fig3} are plotted for HIT elements, and two are for p-n junction SCs. The poorer the surface passivation, the shorter the SCR lifetimes (see the two upper curves); the SCR recombination velocity decreases with the surface passivation quality (three further curves). At first glance, the shorter SCR lifetimes should be found in the p-n junction SCs: the high-temperature operations are used in their manufacturing, accompanied by the impurity gettering in the SCR. However, this is not observed. If the SCR lifetimes had not decreased compared to the SRH lifetimes, then the SCR recombination velocities would have been $\le 0.01$\,cm/s   (see the two lower curves). In this case, the SCR recombination would practically not affect either the efficiency or the effective lifetimes of the excess charge carriers.

\begin{figure}
    \centering \includegraphics[width=1\linewidth]{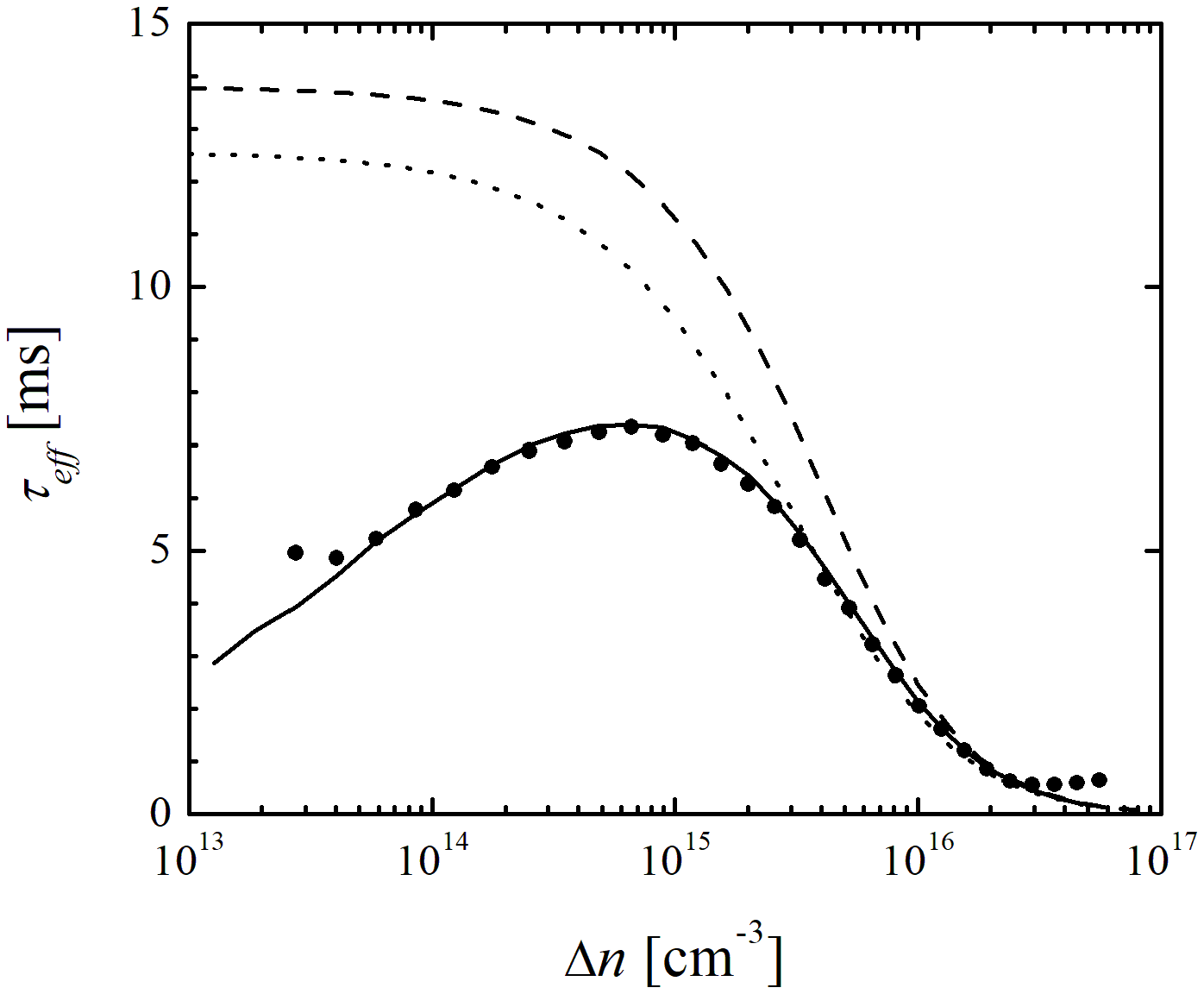}
    \caption{Effective lifetime vs. excess concentration in an SC from the work of Yoshikawa et al.\cite{Yoshikawa17}. Symbols: experimental data. Solid line: theoretical curve. Dashed: The theoretical curve is plotted assuming that the SCR recombination time equals the bulk value. Dotted line: a theoretical curve with SCR recombination neglected and the surface recombination velocity set to 0.16\,cm/s.}
    \label{fig4}
\end{figure}

The theoretical effective lifetimes vs. the excess concentration of charge carriers are calculated both for the SCR lifetime equal to $\tau_{SCR} = 79\,\mu$s determined for this SC (solid curve in Fig.~\ref{fig4}), and also under the assumption that the SCR lifetime is the same as the SRH lifetime in the quasineutral base region, $\tau_{SRH} = 15.5$\,ms (dashed line in Fig.~\ref{fig4}). It can be seen from the figure that both the position and the height of the maximum of the experimental curve are correctly reproduced for $\tau_{SCR} = 79\,\mu$s. The value at the maximum effective lifetime obtained assuming $\tau_{SCR} = \tau_{SRH} = 15.5\,$ms significantly (almost twice) exceeds the maximum value of the effective lifetime in the SCR in this SC.

 At first glance, such a decrease in the lifetime amplitude can be attributed to the surface recombination. But the calculations show that if the SCR recombination is neglected, then it is possible to achieve agreement between the theoretical and the experimental effective lifetime only at relatively high excess carrier concentration $\Delta n > 3\cdot10^{15}\,\text{cm}^{-3}$ (see the dotted curve in Fig.~\ref{fig4}). A similar situation is observed for other SCs analyzed. Thus, if the efficiency of the investigated SCs increases by neglecting the SCR recombination by only about one percent, the situation is completely different for the effective lifetimes. Not to mention the fact that the region falling towards small excess concentrations exists for all SCs and, with very few exceptions, is explained by recombination in the near-surface region of the SC base. Thanks to the SCR recombination, the effective lifetime near the maximum significantly decreases, a fact to which little attention was paid until recently.

 Thus, as the above analysis shows, the SCR recombination not only ensures the decrease of the effective lifetime in the SCs studied at sufficiently small excess concentrations, but also strongly affects its maximum value. Moreover, even in the case of a high surface recombination velocity \cite{Taguchi05} $S_0 = 8.2$\,cm/s, the effect of recombination in the SCR on the photoconversion efficiency exceeds the impact of surface recombination. In this case, the actual photoconversion efficiency of 21.5\,\% increases to 21.9\,\% if surface recombination is neglected, and to 22\,\% if recombination in the SCR is also neglected.

 In the SCs with well-passivated surfaces, surface recombination affects the efficiency much less than the recombination in the SCR. For example, for the SCs reported by Yoshikawa et al.\cite{Yoshikawa17}, neglecting the SCR recombination leads to an increase in efficiency by 1\,\%, while neglecting surface recombination increases the efficiency by 0.12\,\%. Since the SCR recombination has a stronger effect than the surface recombination, which is always taken into account, the SCR recombination must also be considered in SC modeling.

  \begin{figure}
    \centering
    \includegraphics[width=1\linewidth]{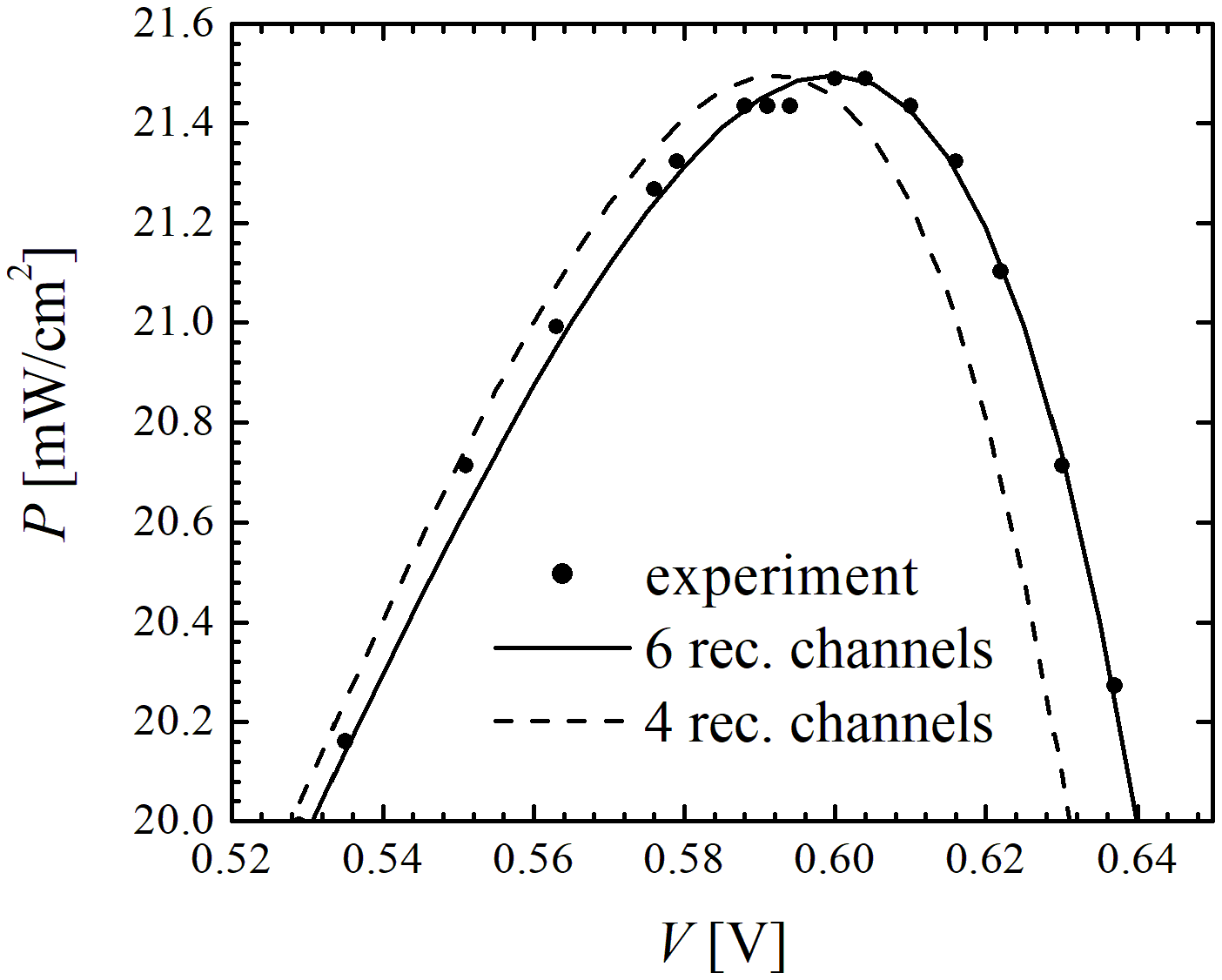}
    \caption{Output power $P$ vs. voltage $V$ across the solar cell, measured experimentally in Taguchi et al. \cite{Taguchi05}  (symbols), and calculated theoretically under the approximation of 6 and 4 recombination channels operative in the SC (solid and dashed lines, respectively). In the four-channel approximation, SRH, radiative, Auger, and surface recombination mechanisms are taken into account. The non-radiative exciton Auger recombination and the SCR recombination in the six-channel calculation are also included.}
    \label{fig5}
\end{figure}

Let us now analyze  the SC output power dependence on the applied voltage. Figure~\ref{fig5} shows this curve for the SCs from Ref.~\onlinecite{Taguchi05} (symbols), compared to the theoretical curves obtained in the approximations of 6 and 4 recombination channels. Within the four-channel model, SRH, radiative, Auger, and surface recombination processes are taken into account. The non-radiative
exciton Auger recombination and the SCR recombination in the six-channel calculation
are also included. It can be seen that near the point of maximum power $V_m = 599$\,mV, the six-channel approximation agrees with the experiment. Although at $V < V_m$, the curves built in these approximations differ only slightly, the discrepancy between them is much greater at $V > V_m$. The locations of the maxima predicted by the two models differ by 7\,mV, which exceeds the error of voltage measurements.

\subsection{Choice of the approximation for Auger recombination rate}
Several approximate formulas exist for the band-to-band Auger recombination rate as a function of the excess charge carrier concentration \cite{Richter12, Black22, Niewelt22, Veith-Wolf18}. 

However, except for a few papers \cite{Lin23}, we are not aware of a comprehensive discussion of the approximation choice for calculating the photoconversion efficiency of highly efficient silicon SCs. Although the effective lifetime changes insignificantly when using different approximations, for sufficiently large excess carrier concentrations, specifically when $\Delta n \geq 10^{16}$\,cm$^{-3}$, the obtained  $\tau(\Delta n)$ curves deviate somewhat more substantially from one another.

A different situation occurs when fitting the illuminated $I-V$ curves and the output power dependence on the applied voltage. While analyzing the experimental illuminated $I-V$ characteristics of the solar cell \cite{Lin23}, we found that the values of the open circuit voltage observed in this work cannot be reproduced theoretically using the approximations of Richter et al. \cite{Richter12}, Black et al. \cite{Black22}, and Niewelt et al. \cite{Niewelt22}. When analyzing the SCs whose surfaces are passivated so well that the surface recombination velocity becomes smaller than 1~cm/s, it becomes important which approximation to use. In the opposite case of surface recombination velocities exceeding 1~cm/s, all approximations for Auger recombination time work equally well. 

\begin{figure}
    \centering \includegraphics[width=1\linewidth]{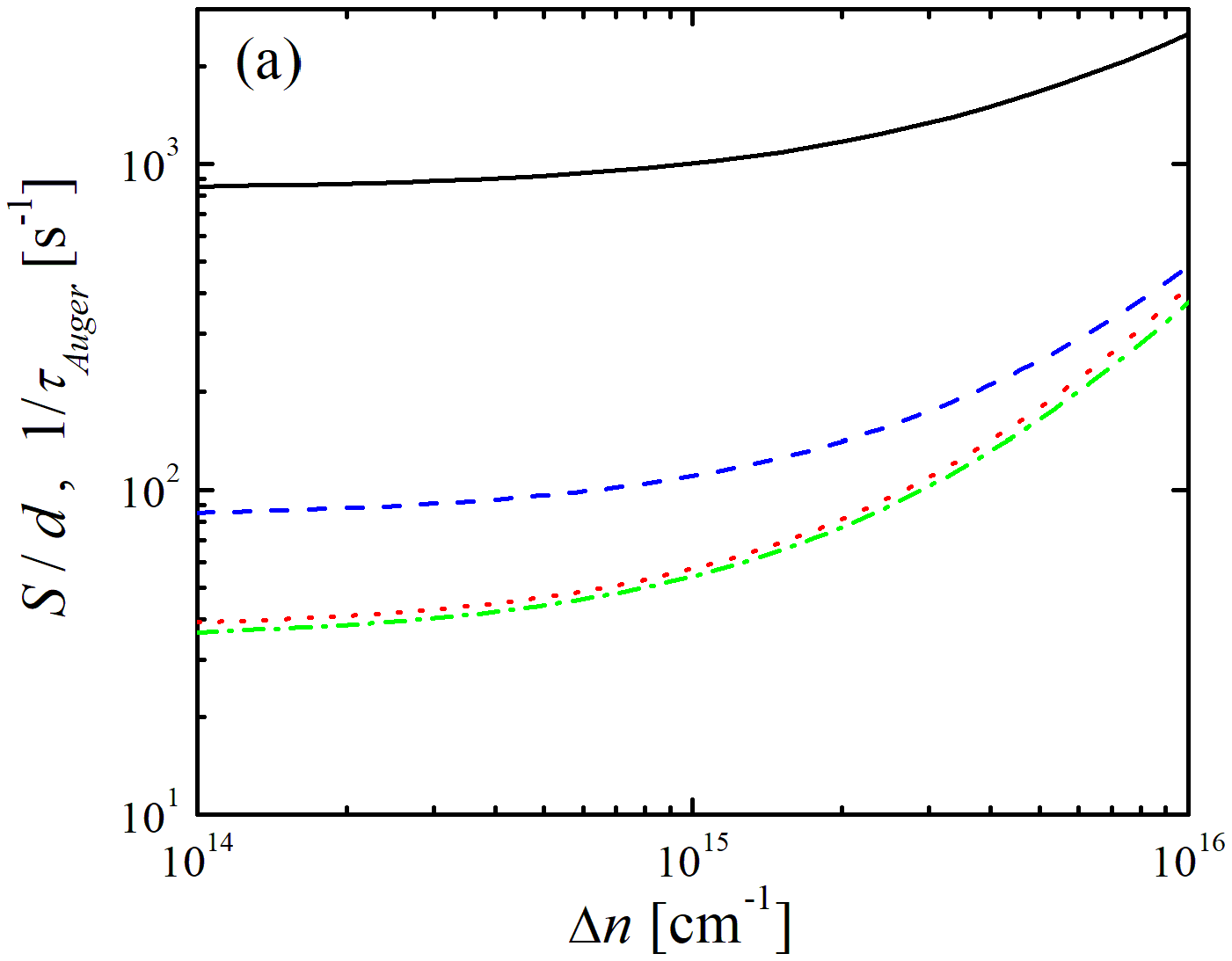}
    \includegraphics[width=1\linewidth]{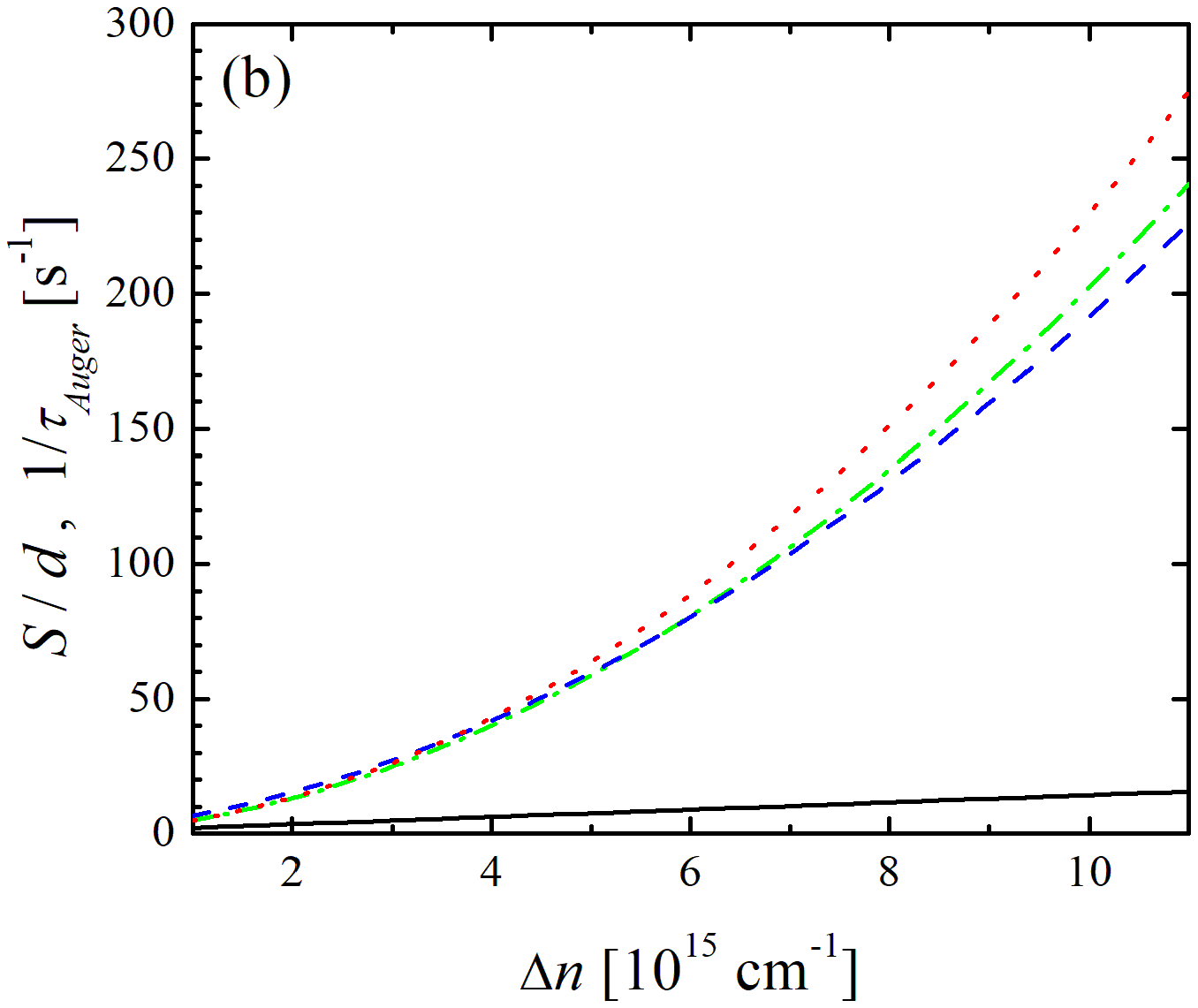}
    \caption{Solid black curves: surface recombination velocity divided by the base thickness from (a) Ref.~\onlinecite{Taguchi05} (doping level $n_0 = 5\cdot10^{15}\,\text{cm}^{-3}$) and (b) Ref.~\onlinecite{Lin23} (doping level $n_0 = 6.5\cdot10^{14}\,\text{cm}^{-3}$). Colored curves: inverse Auger lifetime in various approximations, namely, Richter et al. \cite{Richter13} (blue dashed line), Niewelt et al.\cite{Niewelt22} (red dotted line), and  Veith-Wolf et al. \cite{Veith-Wolf18} (green dash-dotted line).}
    \label{fig6}
\end{figure}

This observation is illustrated by Fig.~\ref{fig6}(a) showing the inverse Auger recombination time $1/\tau_{Auger}$ obtained within different approximations vs. the excess concentration $\Delta n$  for the SC described in Ref.~\onlinecite{Taguchi05}. In this figure, the black solid line shows the surface recombination velocity divided by the base thickness $S/d$, that is, the inverse of the surface recombination time, which significantly exceeds the inverse Auger lifetimes. Therefore, the surface recombination velocity can be found to a good accuracy no matter which approximation is used for the Auger lifetime. 

However, the choice of the approximation matters when the surface recombination velocity is significantly lower than 1\,cm/s. This is the case in the record-efficiency SCs from Ref.~\onlinecite{Lin23}, see Fig.~\ref{fig6}(b), in which $S = 0.017$\,cm/s, with the respective inverse lifetime $S/d$ significantly smaller than the inverse lifetimes of Auger recombination in all three approximations. As can be seen from Fig.~\ref{fig6}(b), the formula from Veith-Wolf et al.\cite{Veith-Wolf18} gives a smaller inverse time than the approximation used by Niewelt et al.\cite{Niewelt22}.  The latter expression allows us to calculate the open circuit voltage in the SC from Ref.~\onlinecite{Lin23} to be about 747\,mV, whereas its experimental value equals 751.4\,mV.

\begin{figure}
    \centering
    \includegraphics[width = 1\linewidth]{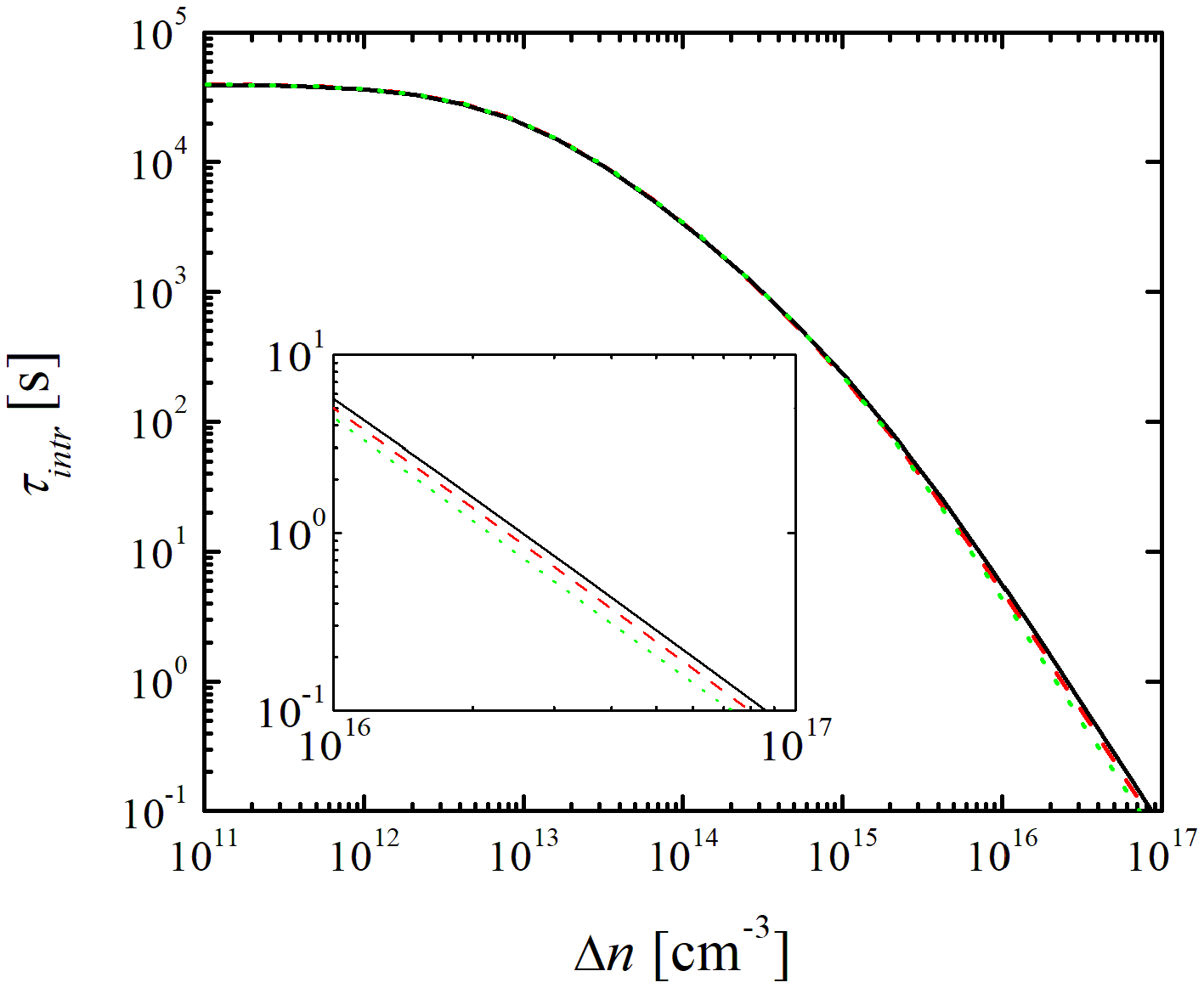}
    \caption{Intrinsic effective (radiative and Auger) lifetime in Si vs. the excess carrier concentration. The inset shows the same plot at $\Delta n$ above $10^{16}\,\text{cm}^{-3}$.}
    \label{fig7}
\end{figure}

Let us further consider the intrinsic recombination limit using for the calculations the parameters from Ref.~\onlinecite{Sachenko20}, namely, $J_{SC} = 0.0434\,\text{A}/\text{cm}^2$, $d = 63.3\,\mu$m, and $n_0 = 10^{13}\,\text{cm}^{-3}$. Figure~\ref{fig7} shows in the double-logarithmic scale the dependence of the effective intrinsic lifetime $\tau_{intr}^{-1} = \tau_{rad}^{-1} + \tau_{Auger}^{-1}$ in silicon on the excess carrier concentration for the three Auger recombination time approximations, proposed by Richter et al.\cite{Richter12}, Veith-Wolf et al.\cite{Veith-Wolf18}, and Nievelt et al.\cite{Niewelt22}, respectively. 

As can be seen from the figure, in the region $\Delta n <10^{15}\,\text{cm}^{-3}$ the curves in the double-logarithmic scale practically coincide. The difference is more pronounced in the semi-logarithmic scale (not shown here). At $\Delta n > 10^{15}\,\text{cm}^{-3}$, the effective lifetime curves calculated using different approximations deviate from each other more substantially (see inset in Fig.~\ref{fig7}). The effective lifetime curve, plotted using the approximation of Richter et al.\cite{Richter12} is most strongly shifted to the right toward higher concentrations. The $\tau_{eff}(\Delta n)$ curve obtained from the formula of Veith-Wolf et al.\cite{Veith-Wolf18} is shifted to the left towards lower concentrations, while the curve constructed using the approximation of Niewelt et al.\cite{Niewelt22} goes even further to the left.  In general, as can be seen from Fig.~\ref{fig7}, the dependencies of the effective lifetime on the excess concentration differ slightly for different approximations for the Auger recombination. Even when $\Delta n = 10^{17}\,\text{cm}^{-3}$, the largest difference between them is 33\,\%, which is not much in these conditions.

To choose the approximation for the Auger recombination rate when modeling the performance of well-passivated SCs, as investigated in Ref.~\onlinecite{Yoshikawa17} and Ref.~\onlinecite{Lin23}, it is useful to analyze their key parameters in the intrinsic recombination limit. As our calculations demonstrate, for the SCs from Ref.~\onlinecite{Yoshikawa17}, the excess concentration in the open-circuit mode is about $1.6\cdot10^{16}\,\text{cm}^{-3}$, and for the SCs from Ref.~\onlinecite{Lin23} it is about $2\cdot10^{16}\,\text{cm}^{-3}$. Table~\ref{table2} shows the key parameters for the SCs with doping levels of $6.5\cdot 10^{14}\,\text{cm}^{-3}$, which corresponds to the case of Ref.~\onlinecite{Yoshikawa17}, and $3.23\cdot 10^{15}\,\text{cm}^{-3}$, which was implemented in Ref.~\onlinecite{Lin23}, in the intrinsic recombination limit using three approximations for the Auger recombination rate from Refs.~\onlinecite{Richter12, Niewelt22, Veith-Wolf18}. We did not consider the approximation from Ref.~\onlinecite{Black22} separately, because we believe that when obtaining the approximation from Ref.~\onlinecite{Niewelt22}, the results presented in Ref.~\onlinecite{Black22} were taken into account.

\begin{table*}[]
    \centering
    \begin{tabular}{| c | c | c | c | c | c |}
    \hline
        SC Ref. &  Doping level, cm$^{-3}$ & $V_{OC}$, mV & $\Delta n_{OC}$, cm$^{-3}$ & $\eta$, \% & Auger approximation\\
        \hline
        \onlinecite{Yoshikawa17} & $6.5\cdot10^{14}$ & \begin{tabular}{@{}c@{}}769.5\\767.5\\764.6 \end{tabular} &
        \begin{tabular}{@{}c@{}}$2.99\cdot10^{16}$\\$2.88\cdot10^{16}$\\$2.70\cdot10^{16}$ \end{tabular} &
        \begin{tabular}{@{}c@{}}29.7\\29.69\\29.6 \end{tabular} &
        \begin{tabular}{@{}c@{}}Richter et al.\cite{Richter12}\\Veith-Wolf et al.\cite{Veith-Wolf18}\\Niewelt et al.\cite{Niewelt22}
        \end{tabular}\\
        \hline

         \onlinecite{Lin23} & $3.23\cdot10^{15}$ & \begin{tabular}{@{}c@{}}766.9\\767.0\\764.4 \end{tabular} &
        \begin{tabular}{@{}c@{}}$2.72\cdot10^{16}$\\$2.72\cdot10^{16}$\\$2.56\cdot10^{16}$ \end{tabular} &
        \begin{tabular}{@{}c@{}}29.1\\29.2\\29.1 \end{tabular} &
        \begin{tabular}{@{}c@{}}Richter et al.\cite{Richter12}\\Veith-Wolf et al.\cite{Veith-Wolf18}\\Niewelt et al.\cite{Niewelt22}
        \end{tabular}\\
        \hline
    \end{tabular}
    \caption{Open-circuit voltage $V_{OC}$, excess carrier concentration in the open-circuit mode $\Delta n_{OC}$, and photoconversion efficiency $\eta$ in the intrinsic recombination limit in the SC of thickness $d = 63.3\,\mu$m and doping level $n_0 = 6.5\cdot10^{14}\,\text{cm}^{-3}$ with three different approximations for the Auger recombination rate from the literature.}
    \label{table2}
\end{table*}

Table~\ref{table2} shows the values of the open-circuit voltage $V_{OC}$, excess concentrations in the open-circuit regime $\Delta n_{OC}$, and photoconversion efficiency $\eta$ obtained for three approximations for Auger recombination rate at two doping levels, which correspond to the ones used in Ref.~\onlinecite{Yoshikawa17} and \onlinecite{Lin23}. 
As can be seen from the table, the experimental values of $\Delta n_{OC}$ are close to those obtained in the intrinsic recombination limit. Therefore, if a particular approximation gives the highest open-circuit voltage and photoconversion efficiency, then it should also apply for the passivated SCs with the surface recombination velocity $S_0 \ll 1\,\text{cm}/\text{s}$. In addition, the expressions for the photoconversion efficiency in the intrinsic recombination limit do not include such important parameters as series resistance. In actual SCs, the series resistance is always present and can significantly affect the photoconversion efficiency. Therefore, a working approximation should be chosen when the series resistance can be ignored, which is one of the defining features of the intrinsic recombination limit.

Table~\ref{table2} shows that when the doping level is $6.5\cdot$ 10$^{14}\,\text{cm}^{-3}$, the highest open-circuit voltage and photoconversion efficiency are reached within the approximation proposed by Richter et al.\cite{Richter17}, and the smallest efficiencies are for the approximation of Niewelt et al.\cite{Niewelt22}.

In contrast, for the doping level of $3.23\cdot 10^{15}\,\text{cm}^{-3}$, the results obtained for the approximations of Richter et al.\cite{Richter17} and Veith-Wolf et al. \cite{Veith-Wolf18} get interchanged, namely, the use of the approximation of Veith-Wolf et al. leads to the highest open-circuit voltage and photoconversion efficiency.  As for the SC from Ref.~\onlinecite{Yoshikawa17}, the best results are obtained with the approximation of Richter et al.\cite{Richter17}, which should be used when modeling its parameters. As for the SCs with substantial surface recombination, any approximation for the Auger recombination time can be used.

\begin{table*}[]
   \centering
  \begin{tabular}{| c | c | c | c | c | c | c | c |}
 \hline
SC Ref. & $n_0$, cm$^{-3}$ & $d, \mu$m & $V_{OC}$, mV & $J_{SC}$, mA/cm$^2$ & $\eta$, \% & $FF$, \% & $R_S, \Omega\cdot\text{cm}^2$ \\
    \hline
    Taguchi et al. \cite{Taguchi08} & $5\cdot10^{15}$ & 98 & 712.4 & 38.6 & 21.5 & 78.23 & 0.69 \\
    \hline
    Sachenko et al. \cite{Sachenko23} & $9\cdot10^{14}$ & 165 & 694 & 40.04 & 21.18 & 77.76 & 1.01 \\
    \hline
    Yoshikawa et al. \cite{Yoshikawa17} & $6.5\cdot10^{14}$ & 200 & 740.3 & 42.5 & 26.63 & 84.64 & 0.101\\
    \hline
    Lin et al.\cite{Lin23} & $3.23\cdot10^{15}$ & 130 & 751.4 & 41.45 & 26.81 & 86.07 & 0.125 \\
    \hline
    Optimized & $7\cdot10^{15}$ & 150 & 745.3 & 42.2 & 27.01 & 85.88 & 0.101\\
    \hline
    \end{tabular}
    \caption{The key parameters of the SCs published in the literature and of the SC from Yoshikawa et al. \cite{Yoshikawa17}, but optimized for its thickness and doping level to maximize the efficiency.}
    \label{table4}
 \end{table*}

 The key characteristics of the SCs we considered, namely doping level $N_d$, base thickness $d$, open circuit voltage $V_{OC}$, short circuit current density $J_{SC}$, efficiency $\eta$, fill factor $FF$, and shunt resistance $R_S$ are listed in Table \ref{table4}. The first four rows of Table~\ref{table4} match those of Table~\ref{table1}, while the last fifth row shows the key parameters of an optimized SC calculated within the formalism that we developed. It should be noted that all the calculated parameter values coincide with the experimental ones up to the third significant figure. Only in the fourth significant figure did the calculated $I-V$ curve fill factors slightly differ from the experimental values.

\begin{figure}
    \centering
    \includegraphics[width = \linewidth]{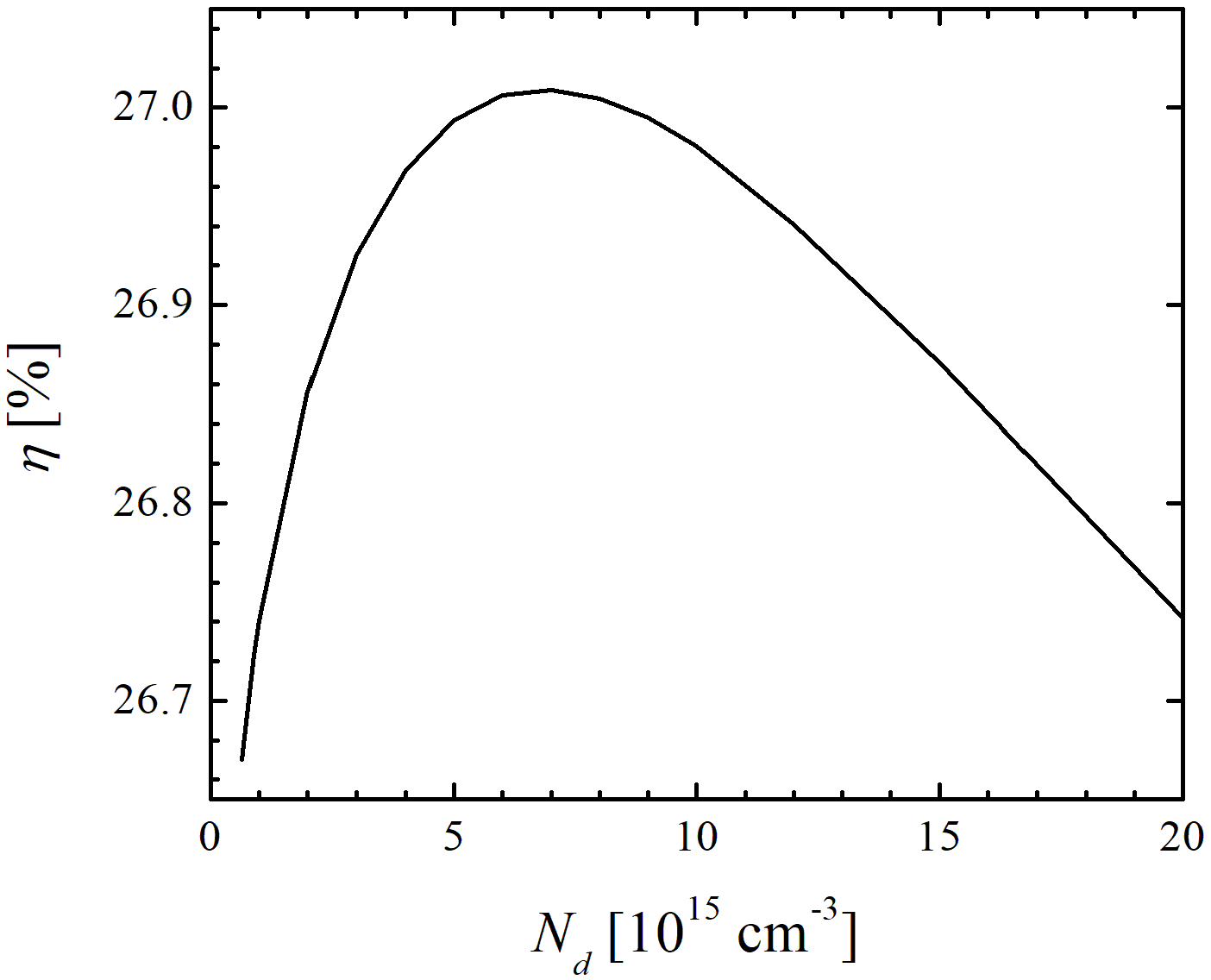}
    \caption{Theoretical photoconversion efficiency vs. the doping level for the SC investigated by Yoshikawa et al. \cite{Yoshikawa17} vs. the doping level.}
    \label{fig8}
\end{figure}

\begin{figure}
    \centering
    \includegraphics[width = \linewidth]{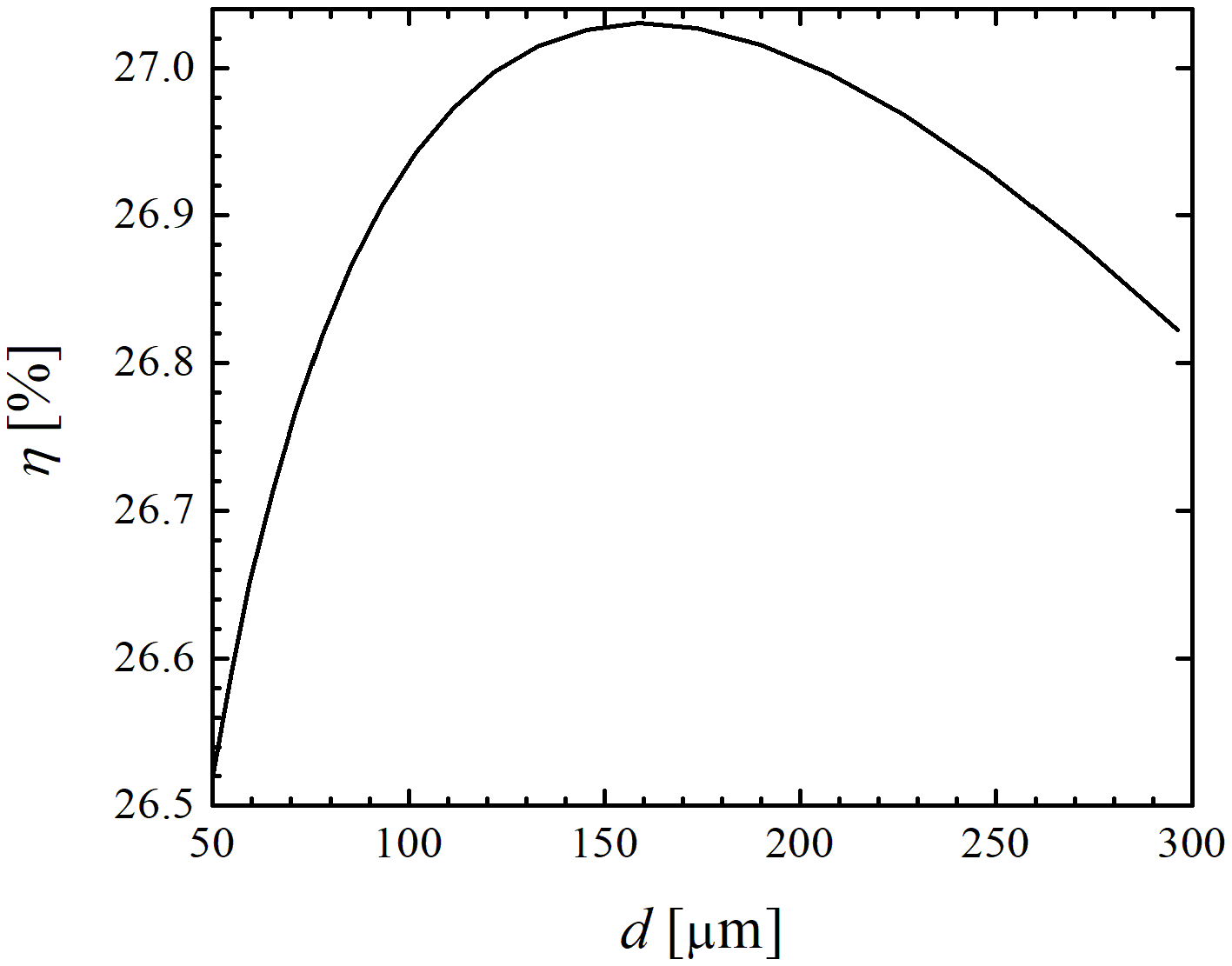}
    \caption{Theoretical photoconversion efficiency vs. the base thickness of the SC fabricated by Yoshikawa et al. \cite{Yoshikawa17}.}
    \label{fig9}
\end{figure}

\subsection{Solar cell optimization}
The theoretical expressions for the efficiency as a function of the doping level and the base thickness can be used to optimize the SCs with respect to these parameters. 
Figure~\ref{fig8} shows the photoconversion efficiency  vs. the doping level for the SC investigated by Yoshikawa et al. \cite{Yoshikawa17}. As can be seen from this curve, the maximum efficiency is reached at the doping level close to $7\cdot10^{15}\,\text{cm}^{-3}$.

Figure~\ref{fig9} shows the dependence of the same SC's efficiency on the base thickness. It is seen that the highest efficiency is reached at the base thickness of 150\,$\mu$m. These results and other parameters are given in the last row of Table~\ref{table4}.

The closeness of the parameters obtained for the optimized sample from Ref.~\onlinecite{Yoshikawa17} and for the SC from Lin et al. \cite{Lin23} (fourth row of Table~\ref{table4}) is striking. Note, in particular, changing the doping level in the optimized sample results in an increase of the open circuit voltage to 745.2\,mV, making it almost equal to the one of the sample from Ref.~\onlinecite{Lin23}. The calculated efficiency of the optimized SC, however, is higher than the efficiency of the record sample \cite{Lin23}, as can be seen from the last two rows of Table~\ref{table4}.

\begin{figure}
    \centering
    \includegraphics[width = \linewidth]{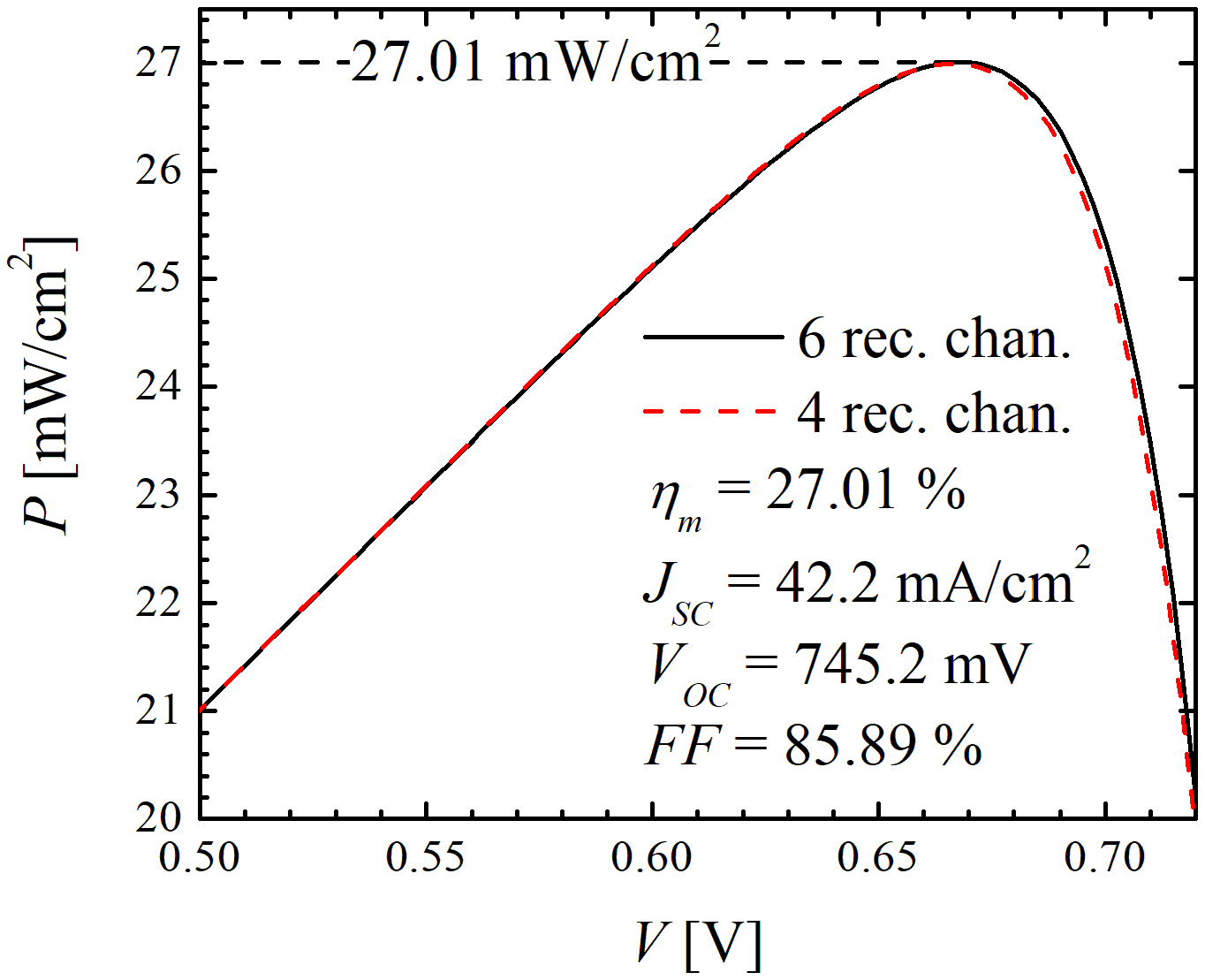}
    \caption{The power vs. voltage curve of an optimized SC from Ref.~\onlinecite{Yoshikawa17} with the recombination rate calculated using six (solid black line) and four (dashed red line) recombination channels.}
    \label{fig10}
\end{figure}

Shown in Fig.~\ref{fig10} is the power vs. voltage curve of an optimized sample using six and four recombination channels. In this case, the difference between the theoretical $I-V$ curves is visually almost imperceptible.
To summarize this part, we can say that the theory developed here allows for the finding of a number of optimal parameters from a simple calculation, whereas experimentally, the same optimization task would require a number of rather expensive operations.

\subsection{Further improvement of the efficiency of silicon SCs}
\begin{figure}
    \centering
    \includegraphics[width = \linewidth]{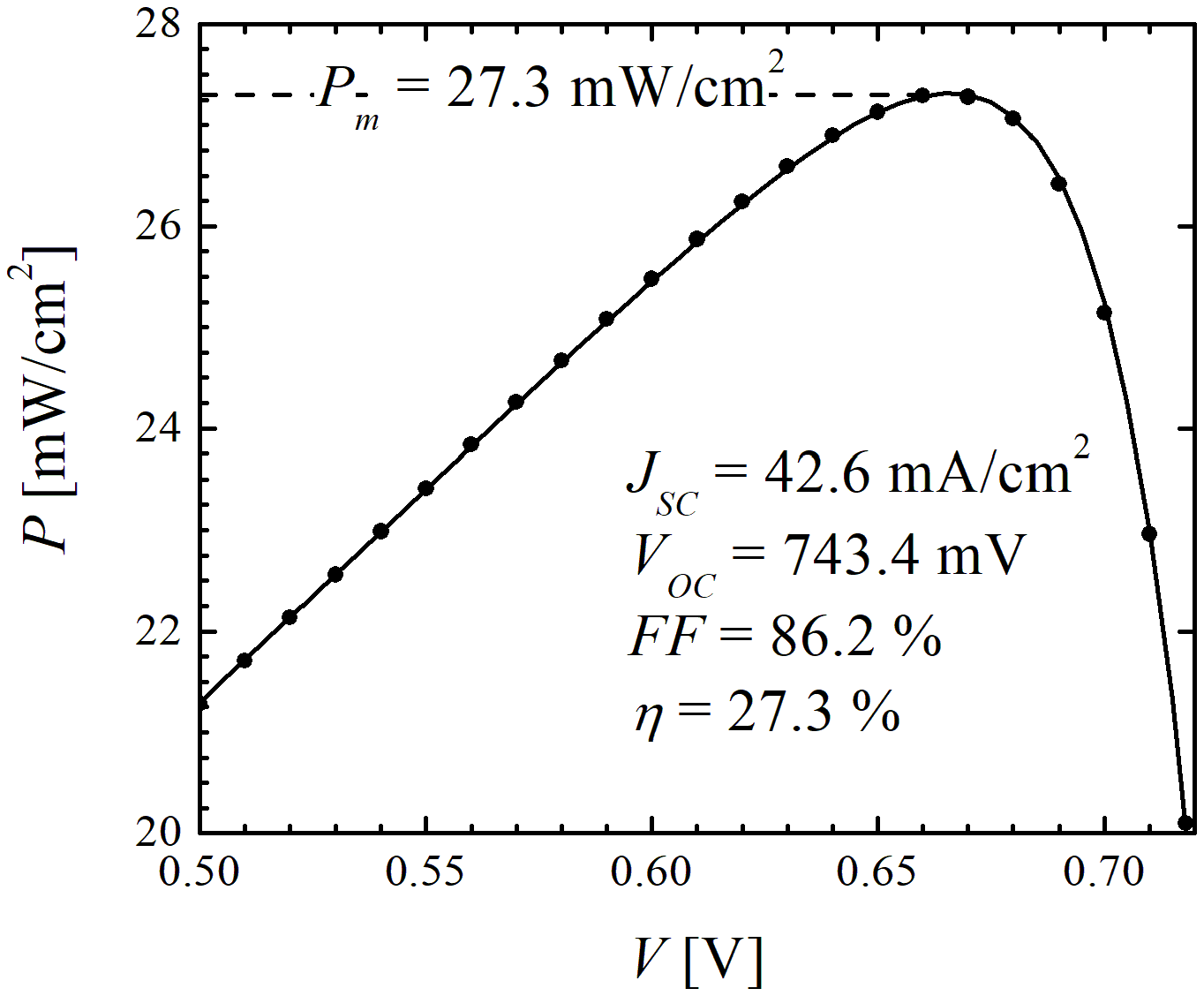}
    \caption{Output power vs. applied voltage for a SC with the record efficiency of 27.3\,\%\cite{Green24}. Symbols: experiment; solid line: theory.}
    \label{fig11}
\end{figure}

Recently, an updated table of high-efficiency SCs been published \cite{Green24}, announcing that a new record of $27.3\,\%$ has been set by the researchers from LONGi company. The article \cite{Green24} also contains the external quantum efficiency and the illuminated $J-V$ curve of this SC. Using the above theory, we calculated the power of this SC on the applied voltage and compared the results with the experiment. The obtained results are shown in Fig.~\ref{fig11}. As can be seen from the figure, the agreement between the theory and the experiment is good.



The results of this analysis are also found in Table~\ref{table12}, which summarizes the parameters of both LONGi SCs \cite{Lin23, Green24}.

In addition, we calculated the efficiency of two hypothetical SCs, in which the SRH lifetime was increased, and it was also assumed that the lifetime in the SCR was equal to the Shockley-Reed-Hall lifetime. These are the third and fourth entries in the tale~\ref{table12}. In the case of the third element, the lifetime is $\tau_{SRH}$ was set equal to 50\,ms, and for the fourth one to 100\,ms. Samples of n-Si with the lifetime of 111\,ms (with $n_0 = 9.43\cdot10^{13}\,\text{cm}^{-3}$) and above 200\,ms (with $n_0 = 6.5\cdot10^{12}\,\text{cm}^{-3}$) were investigated in Refs.~\onlinecite{Black22, Niewelt22}. A strong efficiency increase in these SCs is associated with a decrease of the recombination rate in the SCR. If their lifetime remained equal to 5$\cdot 10^{-2}$ ms, then the efficiency of the third SC would be 27.4\,\%, and the fourth would be 27.5\%. As can be seen, the reduction of the efficiency is significant. This once again emphasizes the need to obtain longer lifetimes for recombination in the SCR.

\begin{table*}[]
\centering
    \begin{tabular}{| c | c | c | c | c | c | c | c | c | c | c |}
   \hline
SC & $J_{SC}$, mA/cm$^2$ & $V_{OC}$, mV & $FF$, \% & $\eta$, \% & $n_0, 10^{15}\,\text{cm}^{-3}$ & $d, \mu$m & $\tau_{SRH}$, ms & $\tau_{SCR}$, ms & $R_S, \Omega\cdot$cm$^2$ & $R_{sh}, \Omega\cdot$cm$^2$\\
\hline
Lin et al. \cite{Lin23} & 41.45 & 751.4 & 86.1 & 26.81 & 3.23 & 130 & 35 & 0.05 & 0.125 & $10^4$\\
\hline
Lin et al. \cite{Green24} & 42.6 & 743.4 & 86.2 & 27.3 & 3.23 & 200 & 35 & 0.05 & 0.15 & $10^5$
\\
\hline
Hypothetical 1 & 42.6 & 744.4 & 86.9 & 27.59 & 3.23 & 200 & 50 & 50 & 0.15 & $10^5$\\
\hline
Hypothetical 2 & 42.6 & 745.2 & 87.5 & 27.72 & 3.23 & 200 & 100 & 100 & 0.15 & $10^5$\\
\hline
\end{tabular}
\caption{Parameters of the SC produced by the LONGi company \cite{Lin23} with the efficiency of 26.81\,\%, the recently reported\cite{Green24} SC with the record efficiency of 27.3\,\%, and two hypothetical SCs with long SRH lifetimes in the base and the SCR.}
\label{table12}
\end{table*}

 \section{Practically achievable maximum efficiency of crystalline silicon solar cells}

Various approaches can be used to assess the practically achievable highest efficiency of the crystalline silicon SCs. 

We propose to combine the parameters of the best-performance SCs to date. Thus, the maximum practically attainable short-circuit current is the one with the smallest coefficient $b$ in Eq.~(\ref{110}), which describes the deviation of absorptance from the Lambertian limit. The smallest value $b = 1.6$ is obtained by fitting the short circuit current of the record-efficiency SCs\cite{Lin23}. Next, the lowest optical losses due to reflection and absorption outside of the SC make up 1\,\%, as can be deduced by fitting experimental data \cite{Yoshikawa17}. These parameters give us the short-circuit current density of $43.09\,\text{mA}/\text{cm}^2$. 

We set the surface recombination velocity to $0.01\,\text{cm}/\text{s}$ and the series resistance\cite{Lin23} to $0.125\,\Omega\,\text{cm}^2$. Furthermore, we take the thickness of the base region to be $110\,\mu$m, the shunt resistance \cite{Yoshikawa17} of $300\,\text{k}\Omega\,\text{cm}^2$ \cite{Yoshikawa17}, the SRH lifetime both in the base region and in the SCR is taken to be $0.05\,\text{s}$, the doping level of $5\cdot 10^{15}\,\text{cm}^{-3}$.

\begin{figure}
    \centering
    \includegraphics[width = \linewidth]{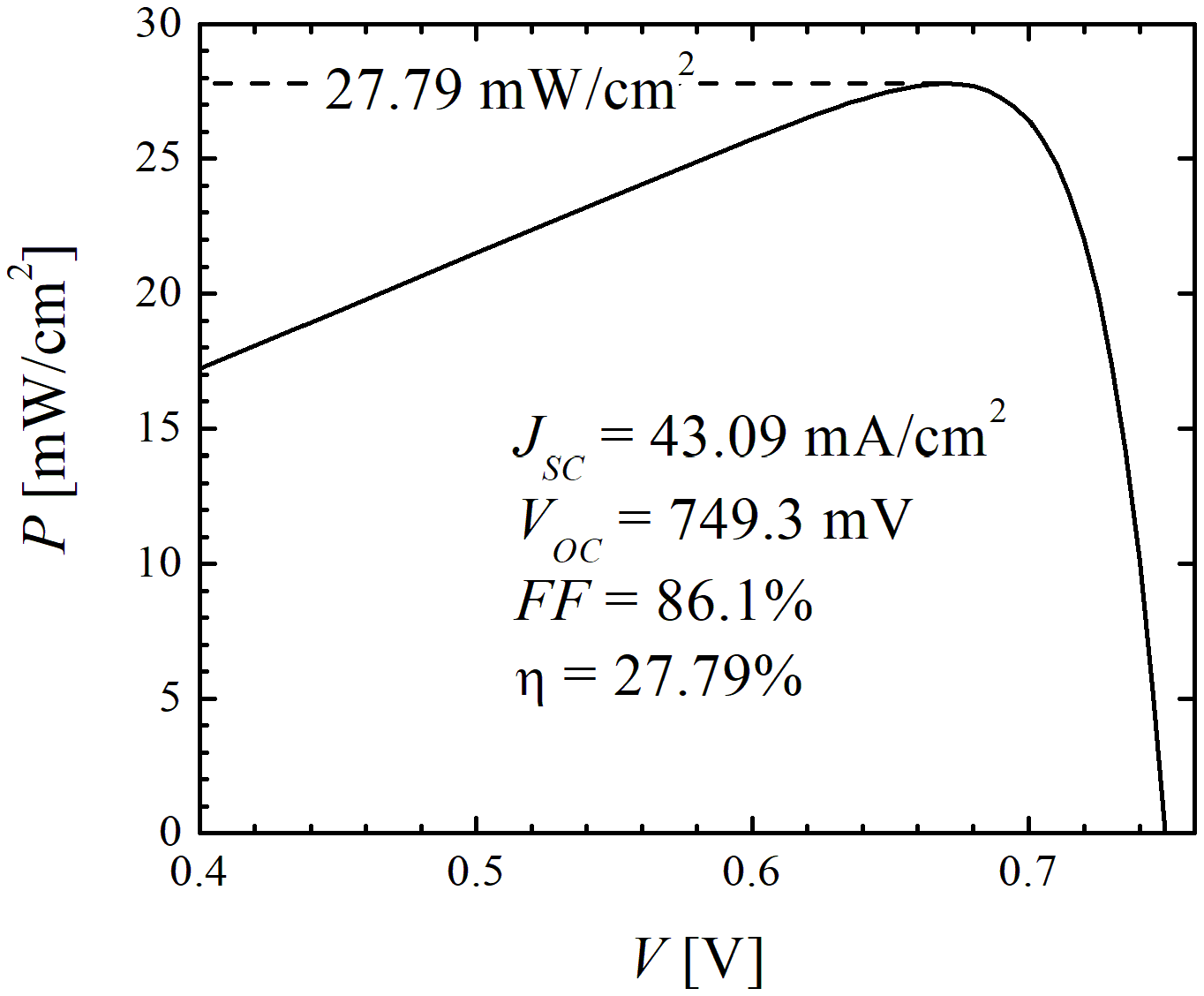}
    \caption{Output power vs. voltage of a hypothetical SC with the practically achievable photoconversion efficiency.}
    \label{fig12}
\end{figure}

The dependence of the power on the applied voltage of the SC constructed in this way is shown in Fig.~\ref{fig12}. The efficiency calculated from them is 27.79\,\%. This value exactly coincides with the achievable efficiency calculated earlier\cite{Su24} for a one-sided SC.

Thus, it follows from the results of our analysis that the practically achievable efficiency of heterojunction SCs is around 27.79\,\%. It can be concluded that the LONGi SC with a photoconversion efficiency of 27.3\,\% \cite{Green24} is very close to the practical limit in terms of photoconversion efficiency, as it is characterized by minimal values of surface recombination velocity and recombination rates in the SCR due to very effective passivation, as well as the best light trapping coefficient $b = 1.6$ compared to other SCs.

Therefore, after reaching the practical limit of photoconversion efficiency, we see further prospects for heterojunction silicon SCs in their use as a bottom part in a tandem with wide-bandgap perovskite semiconductor, despite their somewhat overestimated bandgap compared to the optimal for tandems; this, however, does not lead to a significant decrease in the total efficiency. Organic-inorganic perovskite films, like heterojunction silicon SCs, are characterized by low-temperature manufacturing technology, which is important, and their record efficiency for an area of about 1\,cm$^2$ reached 25.2\,\% \cite{Green24}. Research in the direction of creating perovskite/silicon tandem SCs gives very encouraging results, in particular, monolithic two-contact tandem perovskite/silicon SCs have been reported with the record efficiency values \cite{Aydin24, Green24}, which are 34.2\,\% for an area of 1\,cm$^2$ and 28.6\,\% for an area of 258.14\,cm$^2$.

\section{Challenges to be addressed to reduce  the SCR recombination}

Contrary to the common point of view that the main source of losses in high-efficiency c-Si is surface recombination, the analysis carried out in this work shows that this is not the case. Greater losses are caused by recombination in the SCR, which inevitably exists in every silicon SC, because in each element there is a conductivity inversion, at which the recombination in the SCR is maximal An additional problem is that in practical SCs, the SCR lifetime is significantly shorter than in the quasineutral base region. 

This applies not only to SCs. A similar situation occurs when silicon is passivated by SiN$_x$ or Al$_2$O$_3$ layers\cite{Dauwe04, Veith-Wolf18a}, when a significant charge is built into the dielectric, as discussed earlier in Section~\ref{section3}. The SCR lifetimes must be significantly shorter than in the neutral bulk of the base and are of the order of $0.1-1\,\mu$s\cite{Dauwe04, Veith-Wolf18a}.

From the results presented in this work, it follows that surface passivation, which reduces the surface recombination velocity, also leads to a decrease in the SCR recombination rate. However, there are not enough statistics here to state this for sure. In our opinion, it is worth carrying out targeted research devoted to the analysis of the regularities of the recombination in the SCR and the development of ways to reduce it.

\section{Conclusions}
In this paper, we first briefly review the progress in the research field of
monocrystalline silicon-based photovoltaics. Second, we discuss the semi-analytical
formalism, developed to comprehensively characterize and optimize the highly efficient SCs based on monocrystalline silicon. Third, using the formalism developed we performed analysis and optimized the record-breaking
c-Si SCs.

A numerical parameter is introduced with an analytical expression, which allows for
obtaining accurately reproducing the theoretical wavelength-dependent external quantum efficiency $EQE$ in the long-wavelength absorption spectral
region. The theoretical $EQE (\lambda)$ curves that were obtained are consistent with
the experimental ones.

It has been established that the full characterization of these SCs cannot be carried out without measurements and appropriate processing of the dark $I-V$
characteristics or, alternatively, of the short-circuit current on open-circuit voltage. We proved that it is not possible to reconcile the dependence of the output power on the applied voltage $P(V)$ and the effective lifetime of non-equilibrium charge carriers on the excess concentration with the experiment, without taking into
account the recombination in the space charge region (SCR). 

The procedure for finding the recombination velocity in the SCR using the $J_{SC}(V_{OC})$ or the dark $I-V$ curves is proposed and implemented.

Optimization of the efficiency of silicon SCs depending on the base thickness and the level of doping was carried out. Using the example of specific SCs, it is shown how their photoconversion efficiency can be improved through optimization.

It was established that in the silicon SCs under consideration, the recombination in the SCR reduces the efficiency of photoconversion by approximately one percent, while the effective lifetime due to this effect decreases by several times. It is also shown that in highly efficient silicon SCs, recombination in the SCR affects key parameters stronger than surface recombination.
The results obtained in this work can be used both for the full characterization of highly efficient SCs present today and for the optimization of the key parameters of the next-generation SCs.

\bibliography{refs}

\end{document}